\documentclass[12pt]{article}
\usepackage{setspace}
\usepackage[margin=2cm,nohead]{geometry}
\usepackage{appendix,graphicx}
\usepackage{amsfonts,amssymb,amsmath,mathenv}
\allowdisplaybreaks[3]

\numberwithin{equation}{section}

\newcommand{\nc}{\newcommand}
\nc{\be}{\begin{equation}}
\nc{\ee}{\end{equation}}
\nc{\ba}{\begin{array}}
\nc{\ea}{\end{array}}
\nc{\by}{\begin{eqnarray}}
\nc{\ey}{\end{eqnarray}}
\nc{\la}{\label}
\nc{\ga}{\gamma}
\nc{\mc}{\mathcal}
\nc{\mf}{\mathfrak}
\nc{\mbb}{\mathbb}
\nc{\trm}{\textrm}
\nc{\tbf}{\textbf}
\nc{\mbf}{\mathbf}
\nc{\sig}{\sigma}
\nc{\commut}[2]{\left[ #1{,}\,#2 \right] }

\def\ha{{\textstyle{\frac{1}{2}}}}

\def \adss {$AdS_5 \times S^5$\ }
\def\ci{\cite}
\newcommand{\rf}[1]{(\ref{#1})}
\def \ov {\over}
\def \l {\lambda}
\def\foot{\footnote}

\def \sql {{\sqrt \l}}
\def \m {\mu}
\def \ed {\end{document}}
\def \elk {\ell}
\def \ff {{\rm f}}
\def \aa {{\rm a }}
\def \ads {$AdS_3 \times S^3$\ }

\def \st {{\rm ST}}
\def \pr  {{\rm PRT}}
\def \k {\kappa}\def \s {\sigma} \def \G {\Gamma}

\def \S {{\cal S}}\def \J {{\cal J}}\def \la{\label}
\def \bea {\begin{eqnarray}}
\def \eea {\end{eqnarray}}
\def \be {\begin{equation}}
\def \ee {\end{equation}}
 \def \ta   {\td a}   \def \td {\tilde}
\def \iff {\iffalse}
\def \hh {\mathfrak{h}}

\def \aa  {{\rm a}}

\def \M {{\cal M}}
  \def \ads {$AdS_3 \times S^3$\ }
  
\def \no {\nonumber}

\def \gr {\mathfrak{g}}\def \f {{\rm f}}

\def \ww {{\rm w}}
\def \hSigma {\hat \Sigma}

 \def \ta {\bar \aa}
 \def \o   {\omega}
 
 \def\K{{\rm K}}
 
\begin{document}

\overfullrule=0pt
\parskip=2pt

\headheight=0in \headsep=0in \topmargin=0in \oddsidemargin=0in

\vspace{ -3cm}
\thispagestyle{empty}
\vspace{-1cm}


\rightline{Imperial-TP-YI-2011-01    }

\begin{center}
\vspace{1cm}
{\Large\bf  
Two-loop corrections to partition function of\\ 
\vspace{0.2cm}
 Pohlmeyer-reduced theory  
for ${AdS}_5 \times { S}^5$ superstring 
\vspace{1.2cm}
}

\vspace{.2cm} 
{Y. Iwashita$^{a,}$\footnote{ yukinori.iwashita07@imperial.ac.uk }, 
R. Roiban$^{b,}$\footnote{radu@phys.psu.edu}
  and A.A. Tseytlin$^{a,}$\footnote{Also at Lebedev  Institute, Moscow. tseytlin@imperial.ac.uk}}\\
\vskip 0.4cm
{\em 
$^{a}$ Blackett Laboratory, Imperial College,
London SW7 2AZ, U.K.  \\
\vskip 0.08cm
\vskip 0.08cm $^{b}$ Department of Physics, The Pennsylvania  State University,\\
University Park, PA 16802, USA
}
\end{center}

\vspace{.4cm}

\setcounter{footnote}{0}
\begin{abstract}
Pohlmeyer reduction of \adss  superstring, 
involving solution of  Virasoro conditions in terms of coset current variables, 
leads to a set of equations of motion following 
from  an action containing a  bosonic $Sp(2,2) \times Sp(4)/[SU(2)]^4$ 
gauged WZW  term, an integrable potential  and a fermionic part coupling  bosons from the two
factors. The original  superstring and the
reduced  model are  in direct correspondence at the classical   level 
 but their  relation at the quantum level remains an open question.
As was found earlier,  the one-loop partition functions 
of the two theories  computed  on the   respective classical backgrounds match; 
here we explore the fate of this relation at the two-loop level. We consider the  example 
of the reduced theory solution corresponding to the long folded spinning  string in $AdS$.
The  logarithm of the $AdS_5\times S^5$ superstring partition function computed on the spinning 
string  background  is known to be proportional to the universal scaling function 
which 
depends on the string tension $\sql \ov 2 \pi$. Its ``quantum'' part is 
$f(\l) = a_1 + {1 \ov \sql} a_2 + ...$, 
where the one-loop term is $a_1 = - 3 \ln 2$  and the two-loop term is the negative of the 
Catalan's constant, $a_2=-\K$. We find that the counterpart of $f(\l)$ in the 
reduced theory  is $\ff(k) =\aa_1 + {2 \ov k} \aa_2 + ...$, where $k$ is the coupling  of 
the reduced theory. Here  the one-loop  coefficient is the same as in string
theory, $\aa_1=a_1$,  
while the two-loop one is $\aa_2 = a_2 - {1 \ov 4} (a_1)^2 $. 
Remarkably,  the first Catalan's constant term here matches the string theory result if we 
identify the two couplings as $k= 2 \sql$. 
Nevertheless,  the presence of the additional $(a_1)^2\sim (\ln 2)^2$  term suggests that the 
relation between the two  quantum partition functions (if any) is not a simple equality. 
Similar results are found  in the case of the \ads   string theory  where  
$a_1= - 2 \ln 2$  and  $a_2=0$, while  in the corresponding reduced theory  
$\aa_1=a_1$  and 
 $\aa_2= a_2-{1 \ov 4}(a_1)^2$.
\end{abstract}

\newpage
\tableofcontents

\section{Introduction and summary}

Pohlmeyer reduction  applied  to classical \adss superstring theory  leads \ci{gt1,ms}
to a generalised sine-Gordon type model  described by a particular bosonic 
 $G/H$  gauged WZW theory  with 
an integrable
potential coupled to fermions. Since the Virasoro conditions  are solved in the process of 
the Pohlmeyer reduction, the reduced theory may be viewed as an analog of a 
``light-cone gauge-fixed'' 
version of  the original string theory. 
While  the conformal-gauge 
 \adss  string theory (ST)   and the 
associated  Pohlmeyer-reduced theory (PRT) are closely related  at the classical level, 
PRT has  important  simplifying features being 2d Lorentz invariant and quadratic in
 fermions which have  standard 2d  kinetic terms. When  expanded near the respective 
  vacua, 
 the two theories  are described  by the 
   equivalent sets of  8+8  boson+fermion physical  2d fields. 
  This raises a hope that PRT may be useful  in an attempt to solve the 
 \adss  ST from first principles. 
 
 The relation between the  \adss  ST and PRT is established 
 at the classical level and  involves a transformation from
   coset currents to new fields in a way that solves the 
   conformal gauge  conditions algebraically
   (both  theories originate from the same  
   set of first-order  equations  for the currents).
 The  classical solutions are thus in correspondence  (though  the  
 values of the two actions  on 
 the associated   solutions are, in general,  different). 
 There are  also   close  similarities  between 
  the  associated  tree-level S-matrices \ci{hts}  and the  soliton spectra of the 
   theories  \ci{hm}.
   
 The relation between the \adss  ST and PRT at the quantum level is a priori unclear. 
 Nevertheless, given their classical connection, and the  integrability and UV finiteness 
 of both theories \ci{rtfin}, 
 one may conjecture that the two quantum theories should also  be closely related.
 The precise form  of such a   relation remains to be understood. 
An indication of a  quantum relation is the equality of the one-loop partition functions 
of the two theories computed by expanding  near ``dual'' solutions \ci{hit,iwa}
 \be
 Z_{\rm ST}^{^{(1)}} = Z_{\rm PRT } ^{^{(1)}}  \la{eqpa} \, .
\ee 
  While one may be tempted to 
   view  this  one-loop relation as a consequence  of  the classical equivalence 
   of the  two theories  (suggesting that  determinants  of small fluctuation operators 
    found by perturbing  the classical solutions  should match), 
  it is still 
  a non-trivial test\foot{One may, in principle, construct a pair of classically equivalent
    theories that have different one-loop partition functions.}
     of the  correspondence between the underlying 
   physical degrees of freedom 
   of the two theories.
 
%
The aim of the present paper is to explore possible relations between the two quantum 
partition functions at the two-loop level.
 Since the two-loop  computations  in a non-trivial  background are, in general, very complicated
 here we will  consider  the simplest  string solution -- the infinite spin (scaling) 
 limit of the 
 folded spinning $(S,J)$ string in $AdS_3 \times S^1$ subspace of \adss 
  \ci{ft1,ftt} and the 
 associated  solution of
 the reduced theory. As a further simplification we will eventually 
  consider the limit $J \to 0$
 when the logarithm of the string theory worldsheet partition function is simply 
 proportional to a function of the  coupling constant (i.e. to the universal scaling function of 
  string tension  on the string
 theory  side).

 Our conclusion (in both \adss and \ads  cases that we consider below) 
  appears to be as follows: 
  while the  non-trivial parts of the two two-loop partition functions 
 (coming from the  most complicated   two-loop integrals) 
 appear to be direct correspondence, the reduced theory partition function
 contains an extra two-loop term proportional to the square of the one-loop coefficient. 
 Thus if the two quantum  partition functions are indeed related,  
 this relation  may  be effectively  non-linear. 
 It is  possible also  that the matching of two partition functions 
 may be restored by modifying the PRT action   by  a certain 
  one-loop  counterterm that may be required to maintain 
 its  quantum integrability, i.e. to preserve  certain hidden (super)symmetries. 
 These ideas   remain to be explored.\foot{At the moment we do not have a natural 
 suggestion 
 for a local counterterm that would restore the two-loop  equivalence.}

\subsection{Quantum partition function in string theory}

Let us   first review the known 
 structure of the two-loop string partition in the 
long spinning string background  \ci{ftt,rtt,rt1,rt2,grrtv}.
 The $(S,J)$  spinning string background in the large spin  limit 
  has the following form in  terms of the 
 \adss embedding coordinates  (below $S= \sql \S, \ J= \sql \J$  where $\sql \ov 2 \pi$  is string tension)
\be \la{sj}
\ba{c}
	Y_0+iY_{5}=\cosh ( \elk \sigma) \,e^{i\kappa \tau}\,, ~~~~~~ Y_1+iY_2= 
	\sinh ( \elk \sigma) \,e^{i\kappa \tau}\,, ~~~~~~  Y_{3,4}=0\,,\\	
	X_1+iX_2=e^{i\mu \tau}\,,\ \ \ \ \ \ \ \ \ \ \ \ \ X_{3,4,5,6}=0\,,
\ea  
\ee
where the parameters $\kappa\gg 1 $, $\elk\gg 1 $ and $\mu$ are related by\foot{Note that our notation  here differs from \ci{rt1,rt2}: $\mu_{\text{here}} = \nu_{\text{there}}$, $\ell_{\text{here}} = \mu_{\text{there}}$ and  $(\J/\ln \S)_{\text{here}} = \ell_{\text{there}}$}
\be
\kappa^2= \elk ^2+\mu^2  \ ,\ \ \ \   \ \ \ \  \elk= {1 \ov \pi} \ln \S \gg 1  \ ,\ \ \ \ \ \ \ \ \J= \mu 
	. \label{vi}
\ee 
We will eventually be interested in the limit $\mu\to 0$ when $\k\to \ell$  is the only scale 
in the problem  so that  one may introduce the rescaled worldsheet coordinates 
$\sigma'= \k \s, \ \tau'= \k \tau$ which in the $\k=\elk \to \infty$ limit span the whole 2-plane.
We  shall  define $V_2 = \int  d\tau' d\sigma'=\kappa^2 \bar V_2$   
as the resulting infinite volume factor.  
The logarithm of the 
resulting quantum partition function is given by\foot{Once one includes the contribution of the 
classical action, the full  scaling  function (or ``cusp anomaly'')  is given by 
$\hat f= \sql + f $.} 
 \bea 
&& \G_\st=-\ln Z_\st= {1 \ov 2 \pi} \  f (\l)\  
 V_2 \ , 
  \la{on} \\
&& f (\l)=  
 a_1   + {a_2 \ov \sql } \  + \  O( {{1 \ov (\sql)^2}}) \ , \la{to}  \\
&& 
a_1 = - 3 \ln 2 \ ,\ \ \ \  \ \ \ \ \ \ \  a_2 =  a_{2 B} + a_{2F} = \K -2 \K = - \K  \ . \la{ky}
 \eea
Here  $a_1$ is the one-loop  and $a_2$  is the two-loop 
contribution ($\K$ is the Catalan's constant).
 In $a_2$ we  indicated  separately the part  coming from purely bosonic graphs 
  and graphs  involving fermions. 
The spectrum  of  the string fluctuation modes  \ci{ft1}  includes 
(in the $\m=0$ limit and after rescaling of masses  by $\k$): 
 one $AdS_3$ mode  with $m^2=4$, two $AdS_5$ modes  ``transverse'' to $AdS_3$ 
with $m^2=2$, five  $S^5$ modes with $m^2=0$ and eight fermionic modes with $m^2=1$. 
Contributions  proportional to 
$\K$ originate   from two-loop ``sunset''  graphs with  three propagators that  are 
expressed  in terms of the   following momentum integrals 
\bea 
&& I[m_i^2,m_j^2,m_k^2] \equiv \int { d^2 q_i d^2 q_j d^2 q_k\over (2 \pi)^4}
{ \delta^{(2)} (q_i + q_j + q_k) \ov 
(q^2_i + m_i^2) (q^2_j + m_j^2) (q^2_k + m_k^2)} \ ,  \la{ijk} \\ 
&&  I[4,2,2]={1 \ov (4 \pi)^2 } \K \ , \ \ \ \ \ \   \ \ \ \   
 I[2,1,1]= {2 \ov (4 \pi)^2 } \K  \ . \la{jk} \eea  
Here both  the bosonic $I[4,2,2]$ and the fermionic $I[2,1,1]$ contributions 
involve the ``transverse'' $AdS_5$ modes with $m^2=2$. 
Since such modes are absent in the case of the   $AdS_3 \times S^3$
superstring theory
one expects to find there no Catalan constant contribution. 
Indeed, as we will  show in Appendix B, in this  case 
 \be 
AdS_3 \times S^3: \ \ \ \ \ \ \ \ \ \ \ \ \ \ \ \ \ \ \ 
  \aa_1 = - 2 \ln 2 \ ,\ \ \ \  \ \ \ \ \ \ \  \aa_2 =  0  \ . \la{kyu}
 \ee

\subsection{Quantum partition function in   reduced theory}

Let us now summarize  the  results of  
the corresponding   two-loop  computation in reduced theory described in detail in the 
 main part of this paper. 
 
The Green-Schwarz \adss   superstring theory  is based on the ${F/G}$ supercoset 
 where $F=PSU(2,2|4)$ and\foot{The canonical notation for $Sp(2,2) \subset SU(2,2)$ is 
 $USp(2,2)$.}
   $G=[Sp(2,2) \times Sp(4)]$ 
 with the action having the following symbolic form \ci{MT,B}
\be \la{suu}
I_\st = {\sql \ov 4\pi} \int {\rm STr} \big[J^{(2)}\wedge *J^{(2)} + J^{(1)} \wedge   J^{(3)} ] \ , \ee
  where 
$J^{(2)}$ and $J^{(1),(3)}$ are 
 the bosonic coset  and the fermionic components  of the $PSU(2,2|4)$ current. 
The associated  Pohlmeyer reduced theory 
is given by    a $G/H$  gauged WZW model (with  $G=Sp(2,2)\times Sp(4)$ and $H=[SU(2)]^4$) 
deformed with an integrable
 potential and  coupled to two-dimensional fermions (originating from projections 
 of the fermionic currents $J^{(1),(3)}$). Its  action is, symbolically, \ci{gt1}
 \be \la{jp}
 I_\pr= {k \ov  8\pi} \int d^2 \s\  \  \Big[ L_{gWZW} (g, A)   + {\rm STr} \big( \mu^2 g^{-1} T g T  
  + \Psi  D(A)  \Psi + \mu \Psi g^{-1} \Psi g \big) 
  \Big]
   \ . \ee 
 Here $g \in G$ is related to the bosonic  coset  current, $A_a$  are components of the 
 $H$ gauge field, 
  $T$  is a constant matrix  chosen in the reduction procedure 
whose commutant ($[H,T]=0$) defines the  subgroup $H$ of $G$ 
and $\mu$ is an arbitrary
   mass scale parameter. $\mu^2$  may be 
 interpreted as  a gauge-prescribed
  value of the classical  $AdS_5$ or $S^5$ stress
   tensor in the original string theory (i.e. $\mu$ is playing the role 
 of $p^+$ in the corresponding string light-cone gauge): 
 the simplest ``vacuum'' configuration corresponds 
 to the BMN geodesic with $\mu$ proportional to the   angular momentum~in~$S^5$.   
 
 The coupling constant $k$ of the reduced theory is undetermined by  
  the classical reduction procedure.
 If the quantum string theory and the quantum reduced theory are to be 
  related at all, $k$ should be related to   the string tension or $\sql$.
  Observing that the $\mu$-dependent terms 
 in the reduced theory Lagrangian  \rf{jp}  are exactly equal to the superstring Lagrangian 
 in \rf{suu}
 (with the components of the coset current 
 replaced by its reduced theory values $ J^{(2)}_+ = \mu T, \  J^{(2)}_- =
  \mu g^{-1} T g $, etc.)  one may conjecture that\foot{Note that it 
   is not clear if $k$ is to be quantized in \rf{jp}
   as   the reduced theory   is defined in Minkowski 2-space and is massive.}
 \be     k = 2 \sql    \la{con} \ . \ee
 While the  matching of the one-loop partition functions \rf{eqpa} is not sensitive to the
 values of the two coupling constants 
  (as they    do not enter the determinants of the quadratic 
  fluctuation  operators)  the comparison of  higher-loop quantum corrections 
  crucially depends on a relation like  \rf{con}.

 Our  aim below will be to compute the two-loop correction to the partition function of the PRT 
 defined by \rf{jp} expanded near a solution 
 which is a  counterpart of the 
 long spinning string solution \rf{sj}. In the reduced theory the parameter $\mu$ 
 of the solution in \rf{sj}  becomes identified with the $\mu$ in the 
 PRT action \rf{jp}.  While we shall keep $\mu$ non-zero   at the intermediate stages, 
 to be able to obtain the explicit  two-loop result we will  take the $\mu\to 0$
 limit  in the final expression, i.e. we  will do the 
  two-loop quantum  PRT  computation  for the counterpart of the  long spinning string with $J=0$. 
 In that case  the logarithm of the quantum partition function 
 in the reduced theory has a similar form 
 as  in  string theory (cf. \rf{on},\rf{to}) 
 \bea 
&& \G_\pr=-\ln Z_\pr= {1 \ov 2 \pi} \   \ff (k)\  
 V_2 \ , \la{oni} \\
&&  \ff (k)=  
  \aa_1   + {2  \aa_2 \ov k } \  + \  O( {1 \ov k^2}) \ . \la{toi}  
 \eea
Explicit results  for the  coefficients $ \aa_n$ that  we found 
  are (cf. \rf{ky})
\be \la{kyi}
  \aa_1= - 3 \ln 2 \ , \ \ \ \ \ \ \ \  
    \aa_2 =  \bar\aa_2  +   \tilde \aa_2 \ , 
 \ \ \ \ \  \bar \aa_2  =-\K \ , \ \ \ \ \ 
    \tilde \aa_2  = - {1 \ov 4} (\aa_1)^2 =- {9 \ov 4} (\ln 2)^2   \ . 
 \ee
 The value of the one-loop coefficient $ \aa_1$  matches  the string theory one in \rf{ky}, in
 agreement with \rf{eqpa}. 
  The  Catalan  constant term in $\ta_2$ has exactly the same coefficient as in the string
  partition function in \rf{ky} provided we assume the identification of 
  couplings in \rf{con}.Moreover, the pattern of the bosonic and fermionic contributions
  (i.e. $+\K-2\K=-\K$)  turns out to be  exactly the same as in the 
  string theory expression in \rf{ky}.

   While the mass spectra of the quadratic fluctuation Lagrangians 
   are equivalent, the interaction  vertices 
   could, in principle, generate additional nontrivial contributions in PRT, e.g. proportional 
   to $I[4,4,4]$ in \rf{ijk}, which are not related to the Catalan's constant \ci{rtt}.
   However, all such extra  non-trivial integrals happen not to appear in PRT. 
   We view this and the matching of the Catalan's constant 
   as a strong   indication that the two  
   quantum theories are indeed closely connected.

  At the same time, there is an additional $\tilde \aa_2\sim (\ln 2)^2$ term 
  in the reduced theory two-loop coefficient 
  $ \aa_2$ which is   absent in the string theory  two-loop coefficient $a_2$. 
  To be precise, we did not manage to compute the value of the coefficient 
   of $(\ln 2)^2$ term directly -- we inferred it following a close  analogy with the 
   \ads case where an alternative approach  is available. 
  The    computational procedure we used in the \adss PRT
   (called ``first approach''  below)
  led, in fact,   to an IR divergent result  
   $ 
   \tilde \aa_2=  - \frac{5 }{4 } (\ln 2)^2 - \ln 2 \, \ln m_0$, 
  where $m_0 \to 0$ is an IR   cutoff. We believe this  should be an artifact of 
  our approach in \adss case  in which the  unphysical (non-coset) massless excitations
   were not explicitly
  decoupled.\foot{This approach 
   involves  imposing the $H$ gauge  on  the fluctuation  of the 2d gauge field component, e.g.
   $A_+$, 
  and treating the remaining  bosonic  degrees of freedom, i.e. $g$ and $A_-$ 
  (some of which are unphysical and massless)   on an equal footing. 
  An additional subtlety may be related   to a   particular  way of taking  the  $\mu\to 0$ limit.}
  
  To test the expectation 
  that an alternative     computational   procedure
   that does not involve unphysical  propagating degrees of freedom 
    should  lead to an IR 
  finite result for $ \td \aa_2$   we have  repeated the same two-loop 
  computation in a very similar but simpler setting of the
   reduced theory  for the $AdS_3 \times S^3$ 
  superstring \ci{gt2}. While same  approach as used in \adss case 
   here led  again  to an IR divergent coefficient,  
   $\td \aa_2^{(1)}   = - \frac{2 }{3 } (\ln 2)^2 -\frac{4 }{3 } \ln 2 \, \ln m_0$, 
   the ``second approach'' based on 
   integrating out the 2d gauge fields and gauge-fixing  $g$
   led to a consistent  finite two-loop result  
   \bea  
{\rm AdS_3 \times S^3}: \ \ \ \ \  &&
   \ff(k_3)  =   \aa_1   +  {2  \aa_2 \ov k_3 } \  + \  O( {1 \ov k^2_3}) \ , \\
  \ \ \ \ \  
 && \aa_1 = - 2 \ln 2 \ , \ \ \ \ \ \  \ \ \ \ 
  \aa_2   = -{1 \ov 4} (\aa_1)^2 = - (\ln 2)^2 \ .\la{ads} \eea
  The coupling constants in the \adss  and \ads are related by
  \be k= k_5= 2 k_3   \ . \la{k3} \ee
Once again, the one-loop coefficient here is the same as in \rf{ky} and the absence of the
more complicated  contributions like the Catalan's constant  
 is also consistent  with
the vanishing of the  string theory two-loop coefficient  in \rf{kyu}.
 
 It remains  to be understood if the apparent disagreement of the two-loop 
 coefficients $a_2$ and $\aa_2$ in string and reduced theories by  precisely the 
 square of the one-loop coefficient is still hinting at  some relation  between the
 two  universal scaling functions.


\

The rest of the paper is organized as follows.
In section 2 we shall first review the structure of the reduced theory 
and explain the 
approach to perturbative calculations in it
 based on a field redefinition using
Polyakov-Wiegmann identity and gauge fixing imposed on $A$. 
 We shall present the fluctuation Lagrangian and  list the basic types of
two-loop diagrams  we are going to compute  below. 

In section 3 we  consider the \ads reduced theory. We start with  presenting  the two-loop 
computation using the first  approach explained in section 2 and then consider 
an alternative second  approach based on  integrating out  2d gauge fields
and gauge-fixing imposed on $g$.  
We shall compare the results of the two approaches and suggest  a resolution of 
 the IR  divergence problem of the first approach that should restore   the 
 equivalence  between  the two  approaches. 
 The resulting finite two-loop coefficient is given in  \rf{ads}. 
 
 In section 4 we  present 
 the analogous computation 
  in the \adss  reduced theory.
 We first discuss   the one-loop approximation where 
 the result for the partition function matches the
 string theory result. We then consider the two-loop 
  computation based on the first approach. 
  Using a direct analogy with the \ads  case we 
  propose a modification of the two-loop result 
   that makes it  IR finite. The final  expression for the two-loop 
  coefficient  is given by the  same Catalan's constant term 
  as found in  string  theory  plus an
  additional term  proportional to the square of the one-loop coefficient. 
  
  Appendix A summarizes  our supermatrix notation.  In Appendix B we present the 
  computation of the two-loop 
  universal scaling function  coefficient $a_2$  in the \ads  superstring theory, concluding   
  that it vanishes, i.e.,  in contrast to the \adss string 
  case, it  does not contain 
  the  Catalan's constant term.
  In Appendix C we include some details of the one-loop 
  computation in section 4.1. In Appendix D we  summarize the  computation of the 
  two-loop partition  function  of the reduced \adss theory  in the vacuum case 
  and show that it vanishes.

\section{Perturbative expansion of the Pohlmeyer reduced \\ theory  near a classical
background \label{prandpw}}

In this section we shall briefly 
review the  action  of reduced theory for 
 string theory in ${AdS}_5 \times { S}^5$  and  \ads
  and then consider  its perturbative expansion around a 
  classical configuration.

\subsection{Reduced theory action   } 

The Green-Schwarz action in $AdS_n \times S^n$ spacetime can 
be formulated as a sigma-model action on the supercoset $F/G$ with $F=PSU(1,1|2)
 \times PSU(1,1|2)$  and  $G=SU(1,1) \times SU(2)$ for $n=3$, 
 and   $F=PSU(2,2|4)$  and  $G=Sp(2,2)  \times Sp(4)$ for $n=5$.
The corresponding 
reduced theory 
 is a $G/H$ gauged Wess-Zumino-Witten model 
  with an integrable potential and two-dimensional fermionic fields
\bea \label{dwzwac}
&&	I_\pr=\frac{k}{8 \pi  } \int d^2\sigma\Big[
{\cal L}_{\rm gWZW}+\ \mathrm{STr}\big( \mu^2\,g^{-1}Tg T  \no \\
		&&~~~+\ \Psi_{_L}{T}{D_+\Psi_{_L}}+
		\Psi_{_R}{T}{D_-\Psi_{_R}} +\ \mu\, 
		g^{-1}\Psi_{_L}g\Psi_{_R}\big) \Big]\,,	
\eea
where ${\cal L}_{\rm gWZW}$ is the Lagrangian of the symmetrically gauged  WZW model,
\begin{equation} \label{gauwzw}
	\begin{split}
\int d^2\sigma\ 
{\cal L}_{\rm gWZW}		  &= {\rm STr} \Big[ \frac{1}{2} \int d^2\sigma g^{-1}\partial_+ g g^{-1}\partial_-g  - \frac{1}{6} \int d^3\sigma 
		g^{-1}dg\, g^{-1}dg\, g^{-1}dg \\
		&~~+~\int  d^2 \sigma  \left(  A_+\,
		\partial_- g g^{-1} -
		A_- \,g^{-1}\partial_+ g-g^{-1} A_+ g  A_-  + A_+ A_- \right) \Big] \,.
	\end{split}
\end{equation}
Here $g \in G$ and $A_\pm$ take values in the algebra of $H$  which is 
 $[U(1)]^2$ in the $AdS_3 \times S^3$ case and $[SU(2)]^4$ 
in the ${AdS}_5 \times { S}^5$ case. 
We 
use the notation $\partial _\pm= \partial _\tau \pm \partial_\sigma$, \ 
$D= \partial  + [A, \ ]$.

In general, the normalization of  the $AdS_n \times S^n$ reduced theory 
 action \rf{dwzwac} depends 
on an index of the corresponding matrix 
representation,  with $k$ differing  by 2 in $n=3$ and $n=5$ cases 
 (see \cite{hts}). We will formally use the same normalization in both cases; 
 when comparing the \ads and \adss results   we should   set $k=k_5 = 2 k_3$
 as in \rf{k3}.

The constant matrix $T$ is chosen as    \cite{gt1}
\be \ba{lcl}
	n=3 &:\ \ \ \ \ \ \  & T= \frac{i}{2} {\rm diag }(1,-1,1,-1) \,, \\
	n=5 &:\ \ \ \ \ \ \ & T= \frac{i}{2} {\rm diag }(1,1,-1,-1,1,1,-1,-1) \,.
\ea\ee
Since $[H,T]=0$, the full  action 
is invariant under the $H$ gauge transformations, 
\be\ba{l}\la{hgauge}
~~~~\, 
g\to h^{-1} g\, h \,,\ \ \ \ \ \ \  
A_\pm\to h^{-1}A_\pm h+h^{-1}\partial_\pm  h\ ,
\\ \Psi_{_{R,L}} \to  h ^{-1}\Psi_{_{R,L}}  h 
\,,
 \hspace{20pt}\ h \in H \ . \ea\ee

\subsection{Gauge fixing and parameterization based on the \\
 Polyakov-Wiegmann identity\label{secpwid}}

In general, a bosonic string solution  corresponds to a  bosonic solution 
of the reduced theory given by some  non-trivial background $(g_0, A_{0+}, A_{0-})$. 
To compute the  quantum  partition of the reduced theory on such classical  background 
one  needs  to fix an $H$ gauge. 
It is natural to identify the physical degrees  of freedom 
as corresponding to the coset part  of fluctuations of  $g$, $\delta g\in {\rm alg}(G/H)$, 
while the fluctuations of $g$ along $\hh=$alg($H$)  and $A_+,A_-$ are ``unphysical'';
%
dim$\hh$ of the latter  should be gauge-fixed and the rest integrated out. 
Then there are several possible choices:

 (i) impose  $H$-gauge on the fluctuations of  the gauge field, e.g., on $A_+$;
 
(ii) impose  $H$-gauge on the fluctuations of $g$;

 (iii) impose some  ``mixed''   gauge on the fluctuations of $g$ and $A_\pm$.
 
 \noindent
 In each of these  cases  one may either integrate  out the
 remaining  unphysical fluctuations  from the very beginning or treat them 
 on the same footing with the physical fluctuations of $g$ in the loop expansion. 
   
 For example,  in computing the perturbative S-matrix  near the trivial vacuum 
 $g_0={\bf I}, \ A_{0\pm}=0$ in \ci{hts} the gauge $A_+=0$ was imposed.  Then  
 the constraint following from
 integration over $A_-$ was solved  explicitly by eliminating from the action the 
  unphysical part of $g$ in
 terms of the physical one, ending up 
 with a local action for the  8+8  physical
 massive bosonic+fermionic fluctuations only. 
  In the case of the  reduced theory for the $AdS_3 \times S^3$ string 
  one may impose a gauge on $g$, then integrate out $A_+,A_-$ 
  ending up with a non-linear  action for the 4+4  physical fluctuations. 
  A  ``mixed'' gauge fixing was used 
 in  \cite{hit,iwa} in discussing the one-loop partition function for   fluctuations 
  near a solution corresponding to 
   a string  moving in the $AdS_3 \times S^3$ part of \adss.  
 This gauge led to the decoupling of
  the  unphysical  fluctuations from the physical ones at the level of 
  the quadratic fluctuation action. 
  
  Directly extending each of these approaches to the two-loop level in  the 
   reduced theory for $AdS_5 \times S^5$  string appears to be rather cumbersome. 
  Imposing $ A_+=0$ gauge and then trying to solve the $A_-$-constraint 
   produces a complicated non-local  quartic fluctuation action.  Integrating  out 
   $A_+,A_-$ first  and gauge-fixing $g$ also leads to very involved 
    fluctuation action. It is also unclear how to find a ``mixed''  gauge
    which would ensure the  decoupling of the unphysical   fluctuations 
    beyond the quadratic fluctuation level.
    
  In discussing \adss case in this paper  we shall follow a  different approach  which 
  may be viewed as a version  of (i). It is based on  gauge-fixing $A_+$  combined with a 
  particular field  redefinition of $g$ while  formally 
  keeping the  remaining ``massless'' unphysical fluctuations on the same footing with 
  the ``massive'' physical ones in the two-loop computation. 
  We shall first change the variables from 
  $A_+,A_-$ to $U,\td U$  as  
 \be \label{atou}
A_{+}=U\partial_+ U^{-1}\,,\ \ \ \ \ \ \ \    ~~~A_-=\tilde{U}\partial_- \tilde{U}^{-1} \,,\ \
\ \ \ \ \ U,~\tilde U \in H\ , 
\ee 
  and then use  the Polyakov-Wiegmann  identity to rewrite the PRT  Lagrangian in 
  \rf{dwzwac} as 
  \bea \label{claslag} 
	&&{\cal L}_\pr ={\cal L} (\td g) -{\cal L}_{\rm WZW}(U^{-1}\tilde{U}) \ \ , \\   
	&&{\cal L} (\td g) = {\cal L}_{\rm WZW}(\tilde{g})
	+  {\rm STr}\Big[ \mu^2 \tilde{g}^{-1}T\tilde{g}T 
+   \tilde \Psi _{_L}T\partial_+ \tilde \Psi _{_L}+\tilde \Psi _{_R}T\partial_-
 \tilde \Psi _{_R} + \mu \tilde{g}^{-1} \tilde \Psi _{_L} \tilde{g} \tilde \Psi _{_R} \Big]\,,
 \no
\eea
  where 
  \be \la{ree}
	\tilde{g}= U^{-1}g \, \tilde{U}\,, \ \ \ \ \ ~~~~
	\tilde \Psi _{_L}=U^{-1} \Psi _{_L} U\,,~~~~
	~~~~~~~~~\tilde \Psi _{_R}=\tilde{U}^{-1} \Psi _{_R} \tilde{U} \,.
\ee
 Such a form of the PRT action  was  used  previously 
 in \cite{rtfin} to demonstrate the UV finiteness of the  reduced model. 
 An  advantage of this parametrization  is that the unphysical  degrees of freedom 
 contained in $A_\pm$  are  isolated in the ``ghost-like'' ${\cal L}_{\rm WZW}(U^{-1}\tilde{U})$
 term but one is still  to deal with  the unphysical $\hh$-part of 
 the fluctuations of $\td g$. 
 The action corresponding to  \rf{claslag}  remains  $H$ gauge-invariant 
 under $U^{\prime} =h^{-1}U$, \ $ \tilde U^{\prime}= h^{-1}\tilde U$. This  requires an $H$ gauge
 fixing; one natural  option is to fix (the fluctuation of) $U$ to be trivial. 
 For example, if $U$ has a trivial classical background, after  gauge-fixing $U=1$
  the resulting action will be equivalent  to the one found in \rf{gauwzw} 
 in the gauge $ A_+=0$:    setting $A_-=\tilde{U}\partial_- \tilde{U}^{-1}$
one then  gets ${\cal L} = {\cal L}_{\rm WZW}(g) -
  \tilde{U}\partial_- \tilde{U}^{-1} g^{-1} \partial_+ g + ...$, 
  and finally  $\td U$-part can be decoupled by a redefinition of $g$.

 At the level of  the classical equations  following from \rf{gauwzw} 
 one  can always choose the 
 on-shell gauge $A_+=A_-=0$ \ci{gt1};  in this case only $g$ (and thus also  $\td g$)
 will have a non-trivial background, i.e.  the contribution of the
  path integral over 
 $\tilde{U}$ will be trivial.
  There will still be  ``unphysical'' degrees of freedom 
 contained  in the  fluctuations of $\td g$: in the 
 case of $G=Sp(2,2) \times Sp(4)$ we will   have $10+10$  bosonic fluctuations 
with  $6+6$  corresponding to $H=[SU(2)]^4$ part and $4+4$   being the  ``physical''  coset ones.
The ``unphysical''  degrees of freedom
 should of course effectively decouple (and cancel against 
other  ``ghost'' contributions  and   contribution of determinant 
  of the change of variables \rf{atou}) 
  in the  final expression  for the quantum 
  partition function  but this 
   decoupling  may  not be manifest. 
 
\subsection{Structure of two-loop  quantum corrections\label{stqc}}

Let us now  consider the  expansion 
of the  Lagrangian ${\cal L} (\td g)$ in \rf{claslag} near a classical 
bosonic solution $\td g_0$.
In what follows we shall   omit  tilde on $g$.
Introducing the fluctuations of $g$ taking values in the algebra $\gr $  of $G$
\be\label{ptb} 
	g=g_0 e^\eta = g_0\big(1+\eta +\frac{1}{2} \eta^2 +...\big)  \,, \ \ \ \ \ \ \ \ \ 
	\eta \in \gr \ , 
\ee
we find for the 
 quadratic,  cubic and  quartic terms  in the fluctuation Lagrangian
\bea \label{pw2} 
	&&\mathcal{L}^{(2)}={\rm STr}\Big[ \frac{1}{2} {\cal D}_+ \eta \partial_- \eta 
		-\frac{\mu ^2}{2} \commut{\eta}{g^{-1}_0 Tg_0} \commut{\eta }{T}    \no \\
	&&\ \ \ \ \ \ \ \ \ +  \Psi_{_R} T \partial_- \Psi_{_R} 
	+ \Psi_{_L} T \partial_+ \Psi_{_L}
	 +\mu g^{-1}_0 \Psi_{_L} g_0\Psi_{_R} \Big] 
  \label{pw3}\ ,  \\
&&	\mathcal{L}^{(3)}={\rm STr}\Big[ -\frac{1}{6} 
	\commut{ \eta}{{\cal D}_+ \eta}  \partial_-\eta 
	-\frac{\mu ^2}{6} \commut{\eta}{g^{-1}_0 Tg_0} 
	\commut{\eta}{\commut{\eta }{T}}\no \\
	&&\ \ \  \ \ \ \ \ \ + \mu \left(
	 g_0^{-1}\Psi_{_L} g_0 \eta \Psi_{_R} - \eta g_0^{-1}\Psi_{_L} g_0
	   \Psi_{_R} \right)  \Big]\ , 
\\
 \label{pw4} 
	&&\mathcal{L}^{(4)}={\rm STr}\Big[ \frac{1}{24}
	 \commut{\eta}{\commut{ \eta}{{\cal D}_+ \eta}}  \partial_-\eta 
	+\frac{\mu ^2}{24} \commut{\eta}{\commut{\eta}{g^{-1}_0 Tg_0}}
	\commut{\eta}{ \commut{\eta }{T}}\no \\ 
	&&\ \ \ \ \ \ \ \ \ + \mu \big( \frac{1}{2}g_0^{-1} \Psi_{_L} g_0 \eta^2 
	\Psi_{_R} + \frac{1}{2} \eta^2 g_0^{-1}\Psi_{_L} g_0 
	 \Psi_{_R}-\eta g_0^{-1}\Psi_{_L} g_0 \eta  \Psi_{_R} \big)    \Big] \ , 
\eea
where  ${\cal D}_+ =  \partial _+ \, +\, \commut{\,g^{-1} _0 \! \partial_+ g_0  \,}{\, \,} 
$.

Under the $G/H$  coset decomposition of the algebra $\gr=
 \mathfrak{m} \oplus \mathfrak{h} $   (induced  by $T$,  which selects 
$H\subset G$  such that $[H,T]=0$, see Appendix \ref{appsupcos})  
we have 
\be \la{kw}
\eta = \eta^\parallel + \eta^\perp \ , \ \ 
\ \ \ \ \  \eta^\parallel \in \mathfrak{m}, \ \ \ \ \ \eta^\perp \in \mathfrak{h}\ , \ \ \ \ 
[\eta^\perp, T]=0  \ . \ee 
Here $\eta^\parallel$  describes the  ``physical'' fluctuations.

Our aim will  be to compute the two-loop corrections to the partition function  of this theory in 
the case of a  special background  corresponding to the infinite spin  limit of the folded
 string \rf{sj}. This is a homogeneous background for which the coefficients in the
 fluctuation Lagrangian will be constant; at  the end we will  take the limit  $\mu\to 0$; then 
  there will be only one common scale $\k$ that can be absorbed into the infinite volume
 factor $V_2$  appearing in the logarithm of the partition function or the 
  ``quantum effective action'' \rf{oni}.

\begin{figure}[tb]
 \begin{center}
  \includegraphics[height=3.cm]{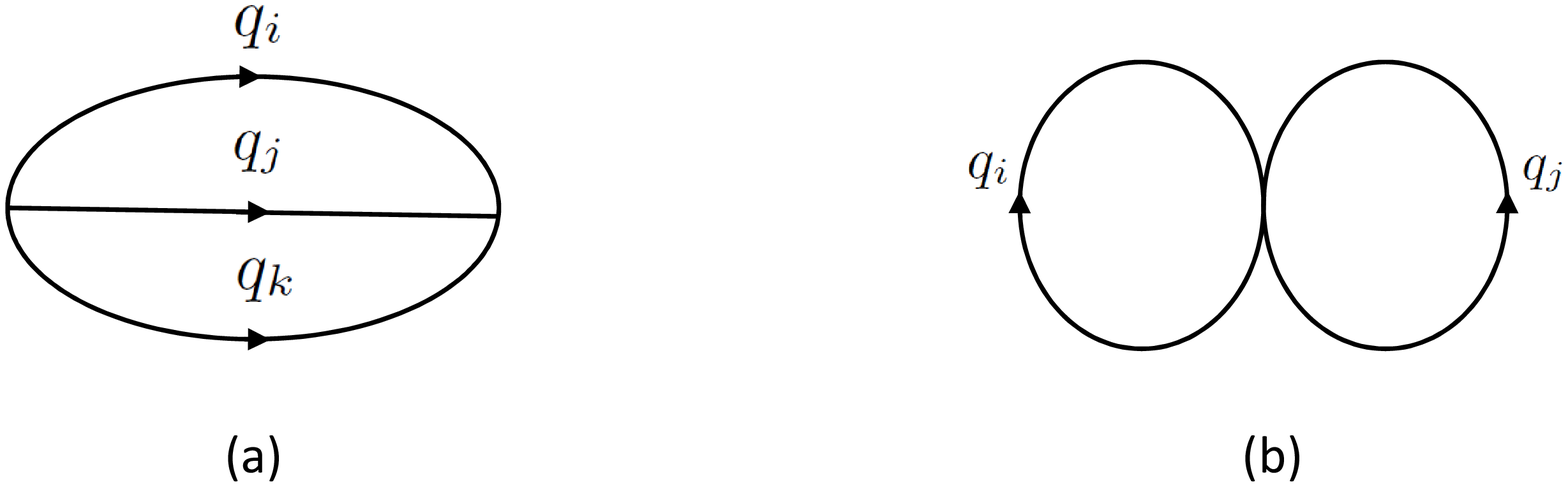}
 \end{center}
 \caption{Bosonic two-loop diagrams.
 In (a)  momentum conservation implies $q_i+q_j+q_k=0$.
  }
 \label{feybos}
\end{figure}

\begin{figure}[t]
 \begin{center}
  \includegraphics[height=3.cm]{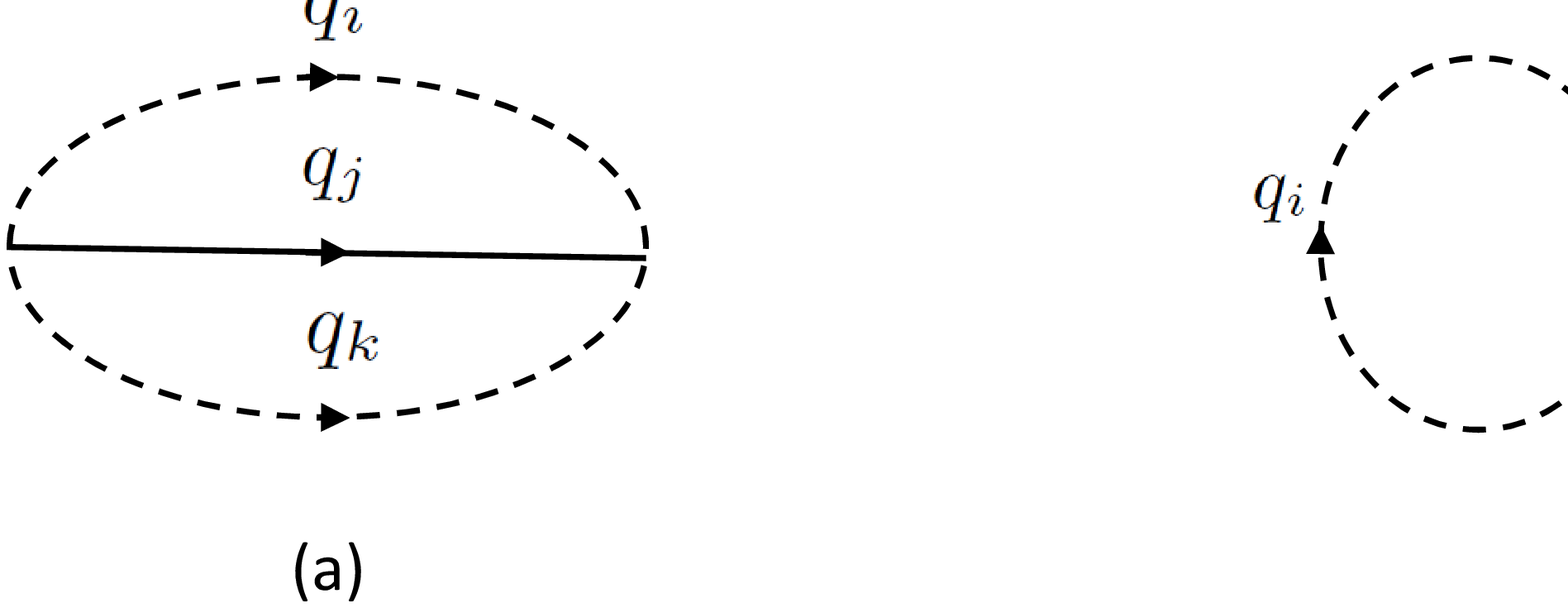}
 \end{center}
 \caption{ Diagrams with bosonic (solid lines) 
 and fermionic (dashed lines)  propagators.
} 
\label{feyfer}
\end{figure}

\begin{figure}[t]
 \begin{center}
  \includegraphics[height=2.cm]{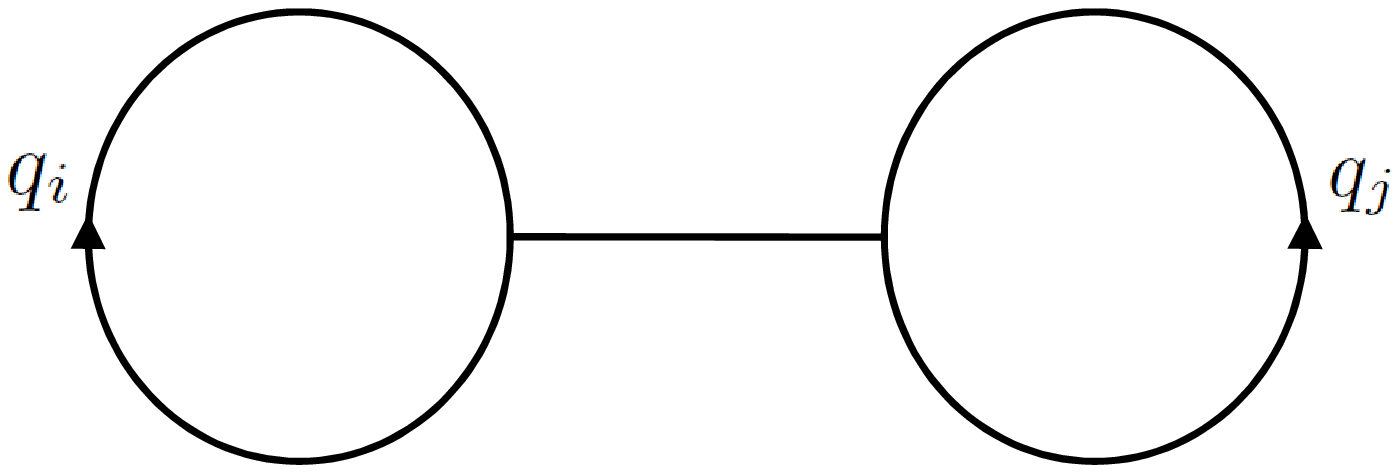}
 \end{center}
 \caption{Tadpole diagrams. The intermediate line is bosonic while loops
  can be either  bosonic or fermionic.}
 \label{feytad}
\end{figure}

The   diagrams  contributing to  two-loop partition functions are shown in  Figures
 \ref{feybos}, \ref{feyfer}, \ref{feytad}.
Let us describe the general form of the  two-loop corrections  on the example of the bosonic
diagrams in  Figure \ref{feybos}. 
 In computing the two-loop  partition function we shall
formally  rotate to euclidean worldsheet time and thus consider the euclidean signature 
propagators.
Writing the bosonic part of the fluctuation Lagrangian in
  \eqref{pw2}, \eqref{pw3}, \eqref{pw4} as 
\be \label{reflag}
	{\cal L}^{(2)}_B + {\cal L}^{(3)}_B + {\cal L}^{(4)}_B =
	 \frac{1}{2} \Phi_{I} \triangle_{IJ} \Phi_{J} +\frac{1}{3!}  V_{IJK} \Phi_{I} \Phi_{J}
	  \Phi_{K}+\frac{1}{4!}  V_{IJKL} \Phi_{I} \Phi_{J} \Phi_{K} \Phi_{L}  + ...\ , 
\ee
where $\Phi_I$ stand for fluctuation field components  and assuming that
 all coefficients are
constant  
one finds that  the one-loop contribution to the logarithm of the 
quantum partition function 
$\G= -\ln Z$  is  given  by 
 $\frac{1}{2} {\rm Tr} \ln \triangle$
 while 1PI part of the two-loop term in  $\G$
   is given   by the sum of the ``sunset'' and
 ``double-bubble'' graph contributions:
\by
	&& \Gamma^{(2)}_{\rm sunset} =-\frac{1}{12}\frac{8 \pi  }{k} V_2 
	\int \frac{d^2 q_i d^2 q_j}{(2\pi )^{2} } V_{IJK} V_{I' J' K'} 
	\triangle^{-1}_{II'}\triangle^{-1}_{JJ'}\triangle^{-1}_{KK'} \label{cint} \,, \\
	&&  \Gamma^{(2)}_{\rm double-bubble} =\frac{1}{8} \frac{8 \pi  }{k} V_2 \int \frac{d^2 
	q_i d^2 q_j}{(2\pi )^{2} } V_{IJKL}  \triangle^{-1}_{IJ}\triangle^{-1}_{KL}\,.
	  \label{qint}
\ey   
Here $-\frac{1}{12}$  and $\frac{1}{8}$  are 
 combinatorial factors, and  $\frac{8 \pi}{k}$ comes from the overall
 factor in front of the action \eqref{dwzwac}. 
 Contributions  of graphs  with fermionic propagators have similar structure 
 (with overall minus sign due to fermionic loop).
 
In general,  individual diagram contributions are gauge-dependent, so some graphs may or may
not appear  depending on the particular gauge fixing.  
In the corresponding 
 two-loop computation in  string theory  non-1PI (``tadpole'') diagrams  did not contribute 
 in the conformal gauge \cite{rtt, rt1}, but were non-vanishing  in the light-cone 
 gauge \cite{grrtv}. 
 In the present case of the reduced theory we will also have non-trivial 
  contributions coming from the  tadpole graphs in Figure \ref{feytad}.
  Moreover, while in lightcone string theory only fermion loops contribute 
  to non-1PI graphs, in the reduced theory both bosonic and fermionic loops yield 
  nontrivial contributions.
%
The intermediate bosonic line connecting the two 
loops there has zero momentum, and 
  the zero-momentum limit may be subtle. Since  
  several physical components of the  propagators (see sections 
\ref{ads3s3} and  \ref{secredads5}) will vanish in that  limit, 
we will set the momentum of the intermediate line to  zero  only 
after doing  the integration in  the two loops.

Let us now  comment on the structure of relevant  momentum integrals.
As in  the string theory computation \cite{rtt,rt1}, we shall assume that all
 power-like UV divergent terms can be regularized away using an analytic regularization
 scheme.\foot{Power divergent terms should cancel out provided 
 all measure factors are properly accounted for.}
 The logarithmic 
 UV divergences should cancel out  \ci{rtfin}
 and we shall verify this below. Our aim will be to compute the finite part of  $\G$. 
Some of the two-loop integrals are expressed  in  terms of   products of  simple 
one-loop integrals,
\be \label{uvandfin2}
 I[m^2] = \int \frac{d^2q}{(2 \pi)^2} \frac{1}{q^2 + m^2} \,.
\ee
It is useful to rewrite  it as 
\be \label{mtoone3}
	I[m^2]=I[1]-\frac{1}{4 \pi} \ln m^2 \,,
\ee
isolating the UV divergent part $I[1]$. 
Double-bubble and tadpole diagram  contributions will be given in terms of 
sums of products $I[m^2_i] I[m^2_j] $. 
The sunset  diagram contributions  are expressed in terms of the 
 following integral  
\be \label{uvandfin3}\ba{l}
\displaystyle I[m_i^2,m_j^2,m_k^2]=\int \frac{d^2 \! q_i \, d^2\! q_j
 \, d^2\! q_k}{(2 \pi)^4} \frac{\delta ^{(2)}(q_i+q_j+q_k)}{(q_i^2 + m^2_i)
 (q_j^2 + m^2_j)(q_k^2 + m^2_k)} \,.
\ea\ee
This integral (already mentioned above in \rf{ijk},\rf{jk})
is UV  finite and also  IR finite for nonzero $m_i$, $m_j$ and $m_k$.  


\section{ Reduced theory for 
 ${AdS}_3 \times {S}^3$  string \label{ads3s3}}

We shall first compute  the  two-loop corrections in the reduced theory 
  corresponding to the 
$AdS_3 \times S^3$  superstring.  The aim will be to compare 
with the superstring result found in Appendix \ref{ads3s3ost}.
In this case $G/H=(SU(1,1) \times SU(2))/[U(1)]^2$
and the action is given by \rf{dwzwac}  with $k=2 k_3$. 
In section 3.1 we shall discuss the
 computation of two-loop corrections in this theory using
the first  approach based on \rf{claslag}  as described in section 2.2.
In section 3.2 we shall consider an alternative approach 
where one imposes  a gauge fixing on $g$ and integrates out $A_\pm$. As
was shown in \cite{gt2},  the resulting model  is the sum of
 the complex sinh-Gordon  and  complex sine-Gordon models
  coupled to  fermions. Here only the  physical degrees of freedom are present 
  and the two-loop computation is straightforward. 

Let  us start with presenting  the classical background $(g_0, A_{0\pm})$ 
in the reduced theory that
corresponds to the  long $(S,J)$ string solution \rf{sj}. 
$g$  is a direct product  of the  ``A'' (i.e. $AdS_3$)  and ``S'' (i.e. $S^3$) parts
corresponding to  $SU(1,1)$ and $SU(2)$, with embedding into $SU(2)$
being trivial. 
If we  choose the basis in $\mathfrak{su}(1,1)$ as  $ R_1=\sigma_1$, $
 R_2=i\sigma_3$ and $ R_3=\sigma_2$  ($\sigma_i$ are the Pauli matrices)
 then   the $G/H$ coset  can be parametrized  by  the Euler angles
 ($\phi$, $\chi$), i.e. assuming vector $H=U(1)$   gauge fixing on $g$ we have  
 $g=\exp \left( \frac{1}{2} \chi  R_2 \right) \exp \left(
  \phi  R_1 \right) \exp \left( \frac{1}{2} \chi  R_2 \right) $. Then,
  a classical  background   corresponding  to a string 
  solution with trivial (BMN vacuum)  ${ S}^3$ part  can be written as 
\bea\label{g3}
&&g=\left(\ba{cc}
	g_{A} & {\bf 0}\\
	{\bf 0} & {\bf 1} \ea
\right)\,, ~~~~~~~~~~
g_{A}=\left(
\begin{array}{cc}
 e^{i \chi } \cosh \phi &  \sinh \phi \\
  \sinh \phi & e^{-i \chi } \cosh \phi
\end{array}
\right) \,, \\
&& 
\label{apmcoth}
A_{\pm}=\left(\ba{cc}
	A_{\pm A} & {\bf 0}\\
	{\bf 0} & {\bf 0} \ea
\right)\,, ~~~
A_{+ A}=-\frac{i}{2}  \partial_+ \chi\ \Big(\ba{cc}
	1 & { 0}\\
	{ 0} & {-1} \ea
\Big)  \,,~~~~~
A_{- A}=\frac{i}{2}  \cosh ^2 \! \phi \, \partial_- \chi\  \Big(\ba{cc}
	1 & { 0}\\
	{ 0} & {-1} \ea
\Big) \ ,\no 
\eea
where the values of $A_\pm$ are determined by solving the classical equations following from 
\rf{dwzwac}.
The fields $\phi$ and $\chi$ appear in the complex sinh-Gordon theory
corresponding to string in  ${AdS}_3\times S^1$  and are related to the global 
${AdS}_3$ string coordinates by 
\begin{equation}\label{redcoth}
\partial_+ Y^P \partial _-Y_P=-\mu ^2 \cosh 2\phi \,, \ \ \ \ \ 
 \epsilon_{QRSP}Y^Q\partial_+Y^R\partial_-Y^S\partial_\pm^2Y^P= 
 4\mu ^3\cosh ^2\!\phi \, \partial_\pm \chi \,.
\end{equation} The  classical reduced theory  background 
 corresponding to the string solution in \rf{sj} 
is
\be \la{jh}
\phi_0 = \ln  \frac{\kappa +\sqrt{\kappa ^2-\mu ^2}}{\mu } \,,~~~~~~~~~~~~~\chi_0 
	=\frac{\mu^2-\kappa ^2}{\mu } \sigma \,.
\ee
Equivalently, the classical PRT   background we are interested in can be  represented  as 
\bea
&&g_0=\left(\ba{cc}
	g_{0A} & {\bf 0}\\
	{\bf 0} & {\bf 1} \ea
\right)\,, ~~~~~~~~~~
g_{0A}=\left(
\begin{array}{cc}
 \frac{ \kappa }{\mu } v_\sigma^* & \frac{\elk}{\mu } \\
 \frac{\elk}{\mu } & \frac{ \kappa }{\mu } v_\sigma
\end{array}
\right) \,, \la{gag}\\
&& A_{0\pm}=\left(\ba{cc}
	A_{0\pm A} & {\bf 0}\\
	{\bf 0} & {\bf 0} \ea
\right)\,, ~~~~~~~~~~
A_{0\pm A}=\frac{i \kappa ^2}{2 \mu }  \left(\ba{cc}
	1 & { 0}\\
	{ 0} & {-1} \ea
\right)\,,     \\ 
&&\label{defvtvs} v_\sigma \equiv  e^{\frac{i
	 \elk^2 \sigma}{\mu }} \,, \ \ \ \ \ \ \ \ \ 
	 v_\tau \equiv  e^{\frac{i \kappa ^2 \tau }{ \mu }} \,,
	~~~~~~~~~~~~~~~
	 \  w \equiv  v_\tau v_\sigma   \ . \la{hag}
	 \eea
Here we  introduced also the  functions $v_\tau$ and $w $ that will be 
often used below. The value of the classical reduced theory action 
on this background is
\be
\Gamma^{(0)}=I_\pr= \frac{k_3}{4\pi}V_2\,\frac{\mu^4-\kappa^4}{\mu^2} \ .
\ee

\subsection{Approach based on   PW identity and gauge-fixing   $A$
\label{ads3s3gwzw}}

The above classical solution written in terms of  the variables 
$U, \td U$ and $\td g$ in \rf{atou},\rf{ree}  becomes
\bea
&& \label{utu3}
U_0=\left( \ba{cc}
u & {\bf 0}\\
{\bf 0} & {\bf 1}\ea
\right) \,,~~~~~
\tilde U_0=\left( \ba{cc}
\tilde u & {\bf 0}\\
{\bf 0} & {\bf 1} \ea
\right) \,, \  \ \ \ \ \ \ \ \ 
u=\tilde u=\left(
\begin{array}{cc}
 v_\tau^{* 1/2} & 0 \\
 0 &  v_\tau^{1/2}
\end{array}
\right)\,,  \\
&& 
 \label{g3pw}
   \tilde g_0=U^{-1}_0 g_0 \td U_0 = \left(\ba{cc}
	  \tilde g_{0A} & {\bf 0} \\
	{\bf 0} & {\bf 1} \ea
\right)\,, ~~~~~~~~~~~~~~~
  \tilde g_{0A}=\left(
\begin{array}{cc}
 \frac{ \kappa }{\mu } v_\sigma^* & \frac{ \elk}{\mu } v_\tau \\
 \frac{ \elk}{\mu } v_\tau^* & \frac{ \kappa }{\mu } v_\sigma
\end{array}
\right)\,.
\eea
We shall now  discuss the computation of quantum corrections
on this background using the approach described in sections 2.2 and 2.3, 
i.e.~using the fluctuation Lagrangian in \eqref{pw2}, \eqref{pw3}, \eqref{pw4}.

\subsubsection{One-loop contribution}


Using  the parametrization described in Appendix \ref{appsupcos}, let us  introduce 
the  following bosonic fluctuation fields in \rf{kw}:  
\bea
&& \label{etapr3}
	\eta^\parallel =\left(
	\ba{cc}
	\eta^\parallel _A & 0 \\
	0& \eta^\parallel _S
	\ea \right) \,,\ \ \ \ \ \ \ \ 	
	\eta^\perp =\left(
	\ba{cc}
	\eta^\perp _A & 0 \\
	0& \eta^\perp _S
	\ea \right) \,,\ \\
&&	
\eta^\parallel _A =\left(
\begin{array}{cc}
 0 &  w (a_1+i a_2) \\
 w^* (a_1-i a_2) & 0
\end{array}
\right)\,, ~~~~~
	\eta^\parallel _S= \left(
\begin{array}{cc}
 0 &  b_1 +i b_2 \\
   b_1-i b_2 & 0
\end{array}
\right) \ ,\\
&&
\eta^\perp _A= \left(
	\ba{cc}
	i c & 0 \\
	0& -i c
	\ea \right) \, , ~~~~~
\eta^\perp _S= \left(
	\ba{cc}
	i d & 0 \\
	0& -id
	\ea \right) \,. 
\eea
The fields $a_1,a_2,b_1,b_2$  represent the physical (coset) fluctuations while 
$c$ and $d$ are ``unphysical'' ones. The factors $w=v_\tau v_\sigma$ are introduced 
so that to make the coefficients in the resulting fluctuation Lagrangian constant. 
Then the ``A'' and ``S''  parts of the quadratic fluctuation Lagrangian   \eqref{pw2}
are found to be (the action is $I={ k_3 \ov 4 \pi}  \int d^2 \sigma \ {\cal L}$)
\bea \label{ads3s3total}
&&	{\cal L}^{(2)}={\cal L}^{(2)}_{{\rm A}}+{\cal L}^{(2)}_{{\rm S}} \,,
\\
&&\label{ads3l}
  {\cal L}^{(2)}_{{\rm A}} =   \sum _{i=1,2}  \partial_+ a_i \partial_- a_i + 2 \left( 
  \mu \partial_+a_2 + M_2 \partial_-a_2 \right) a_1  - \partial_+ c \, \partial_- c - 4M_1
  a_1
   \partial_- c  \,,
\\ &&  \label{bbdef3}
M_1=\frac{\kappa \sqrt{\kappa^2 -\mu^2}}{\mu^2} \,,~~~~~M_2=\frac{  2\kappa^2-
\mu^2}{\mu }\ , \\
&&  \label{s3l}
	{\cal L}_{{\rm S}}^{(2)}= \sum_{i=1,2} \left(  \partial _+ b_i \partial _- b_i -
	 \mu^2 b_i^2 \right) +  \partial _+ d \partial _- d \,.
\eea
The spectrum of massive physical fluctuations 
is exactly the same as in the corresponding string theory \ci{ftt}:
 \eqref{ads3l} describes 
 two physical fluctuations with frequencies 
$	\sqrt{n^2+2 \kappa ^2\pm 2 \sqrt{\kappa ^4+n^2 \mu ^2}} \,,$
(where $n$ is spatial momentum number on worldsheet circle) 
while \eqref{s3l} describes 
two physical fields whose characteristic frequencies are 
$	\sqrt{n^2 + \mu^2 }\,.$
The one-loop partition function  following from 
\eqref{ads3l} and \eqref{s3l}   
\be \label{detbos3pre}
	\Big( \det \left( \partial_+ \partial_- \right)^2 \det\left( \partial_+ \partial_-
	 +\mu ^2\right)^2\det \left[ \partial_+^2 \partial_-^2 +2 \partial_+ \partial_-  \left(2
	  \kappa ^2-\mu ^2\right)+\left( \partial_+^2 + \partial_-^2 \right) \mu ^2\right]
	   \Big)^{-1/2} \,,  
\ee
differs 
  from the bosonic part of the string theory result only by the unphysical 
  massless field contribution $[\det \left( \partial_+ \partial_- \right)]^{-1}$. The latter 
   is canceled out  once we account for (i) 
 the Jacobian of  the transformation \eqref{atou},
 and (ii) the contribution of the $U^{-1} \td U$ dependent   WZW  term in \eqref{claslag}.
 The latter   gives only massless contribution since 
 according to \rf{utu3}  $U^{-1} \td U$ has trivial background and we may gauge-fix the 
 fluctuation of $U$ to be zero (which corresponds to the  $\delta A_+=0$ gauge).
 
 The fermionic fluctuations can be parametrized as follows: 
\bea
	&& \Psi_{_R}=\left(
	\begin{array}{cc}
		0& \mathfrak{X}_R \\
		\mathfrak{Y}_R& 0 
	\end{array}
	\right) \,, ~~~~~~~~~~~~
	\Psi_{_L}=\left(
	\begin{array}{cc}
		0& \mathfrak{X}_L  \\
		\mathfrak{Y}_L& 0 
	\end{array}
	\right) \,,\\ 
&& \label{fxr3}
	\mathfrak{X}_R=\left(
\begin{array}{cc}
 0 & (\alpha_1 +i \alpha_2 ) t_{_{1+}}  \\
  (\alpha_3+i \alpha_4) t_{_{2+}}^* & 0
\end{array}
	\right)\,, \ \ \ \ 
	\mathfrak{Y}_R=\left(
\begin{array}{cc}
 0 & (-i \alpha_3-\alpha_4)  t_{_{2+}}  \\
 (i \alpha_1 +\alpha_2 )t_{_{1+}}^*  & 0
\end{array}
	\right) \no \\
	&&
	\mf{X}_L=\left(
\begin{array}{cc}
 0 &  (\beta_1 +i \beta_2) t_{_{1-}} \\
 (\beta_3+i \beta_4) t_{_{2-}}^*  & 0
\end{array}
	\right)\,,   \ \ \ \ 
	\mf{Y}_L=\left(
\begin{array}{cc}
 0 & (-i \beta_3-\beta_4) t_{_{2-}}  \\
  (i \beta_1+\beta_2) t_{_{1-}}^* & 0
\end{array}
	\right)\,  \label{fyl3}  \no  \\
	&& 
	\label{deftpm}
	t_{_{1\pm}}\equiv e ^{i \frac{\elk ^2 (\tau  \pm \sigma )}{2 \mu }} \,,~~~~~
	t_{_{2\pm}}\equiv e ^{i\frac{\left(\kappa ^2+\mu ^2\right) \tau  \pm \elk^2 
	\sigma}{2 \mu } } \,,
\eea
where  the component fields $\alpha_k, \beta_k$ are real Grassmann. 
The rescaling factors 
$t_{_{1\pm}}$, $t_{_{2\pm}}$ are introduced 
to  make the  coefficients in the  resulting fermionic part of the 
quadratic  fluctuation 
 Lagrangian \eqref{pw2} constant: 
\bea \label{ferpw3} 
&&\mathcal{L}^{(2)}_{\rm F}=\sum\limits_{i=1}^4 \left( \alpha_i \partial_-\alpha_i +\beta_i 
\partial_+\beta_i\right) 
+2 \mu \left( \alpha_3\alpha_4 + \beta_3\beta_4 \right) 
+2 \kappa \left(  \alpha_1\beta_2 -\alpha_2\beta_1   -\alpha_3\beta_4 +\alpha_4\beta_3 
   \right)
 \,.
\eea
The resulting fermionic characteristic frequencies  are:   
$2 \times \sqrt{n^2 + \kappa^2},\  
	\sqrt{n^2 + \kappa^2} +  \mu , \ 
	\sqrt{n^2 + \kappa^2} - \mu $. 
	These are equivalent to the string theory fluctuation spectrum 
	with $\pm \mu$ shifts being due to an overall $\tau$-dependent 
	 rotation of the fluctuations (these cancel out in the  sum over frequencies or in
	 the resulting functional  determinant). Indeed,  
the fermionic  contribution to the one-loop 
partition function  following from \rf{ferpw3} is 
\be\label{detpwf3}
	\big[\det   \left( \partial_+ \partial_- +\kappa ^2\right)\big]^2\
	\det  \left[ \partial_+^2 
	\partial_-^2 +2\kappa ^2 \partial_+ \partial_- +\mu ^2\left( \partial_+^2 + \partial_-^2
	 \right) +\left(\kappa ^2-\mu ^2\right)^2 \right] \,.
\ee
Here the  determinant of the 4-th order operator  can be factorized  as follows:
\bea
&& \det \Big( \left[ \partial_+ \partial_- +i \mu \left( 
	\partial_+ + \partial_- \right) -\mu ^2 +\kappa ^2 \right] \left[ 
	\partial_+ \partial_- -i \mu \left( \partial_+ + \partial_- \right) 
	-\mu ^2 +\kappa ^2 \right] \Big) \no \\
&& = 
	\det \left[ e^{-i \mu \tau} \left( \partial_+ \partial_-  +\kappa ^2
	 \right) e^{i \mu \tau } \right]\  \det \left[ e^{i \mu \tau} \left( 
	 \partial_+ \partial_- +\kappa ^2 \right) e^{-i \mu \tau } \right] \,. \la{shi}
\eea
and thus the fermionic one-loop contribution is equivalent 
to the string theory one: $[\det( \partial_+ \partial_-  +\kappa ^2)]^4$.
Combining the  bosonic and fermionic  contributions  we find 
in the limit $\mu\to 0$ the following expression for the 
 one-loop correction to the effective action ($V_2 = \kappa^2 \bar V_2$) 
\bea
&&\G^{(1)} = {1 \ov 2}  \bar V_2 \int { d^2 q \ov (2 \pi)^2 }\big[ \ln (q^2 + 4 \kappa^2)
 + 3 \ln q^2  - 4 \ln (q^2 +  \kappa^2) \big]\no\\ &&\ \ \ \ \ 
  = 2 \kappa^2 \bar V_2 \big(I[4]-I[1]\big) 
  =  { 1 \ov 2 \pi} ( -2 \ln 2 )   V_2 \ , \la{jok}
  \eea
where the first equality  is proved by  differentiating  over $\kappa^2$. 
Thus  the resulting one-loop  coefficient in \rf{toi} 
	is  given by the same value \rf{ky} as in the $AdS_3 \times S^3$ 
	string theory:  $ \aa_1= a_1=- 2 \ln 2$.


Let us note that  in the above approach based on \rf{claslag} 
the limit $\mu\to 0$, though regular at the level of the quantum partition function, 
 appears to be singular at the level of the 
 fluctuation Lagrangian ($M_1,M_2$ in  \eqref{bbdef3}  blow up). 
This may be attributed to a special nature of the field redefinition/gauge choice we  used.
Indeed, in the ``mixed''  gauge approach used in 
 \cite{iwa} (where unphysical fluctuations were explicitly
decoupled from physical ones) 
 it was found  that this  limit is well-defined in the fluctuation
 Lagrangian. However, this  ``decoupling'' gauge does not appear to have a useful 
  extension beyond the quadratic fluctuation level.


\subsubsection{Two-loop contribution}

As we have seen above, we have $2+2$ physical and $1+1$ unphysical 
 bosonic fluctuations  which are coupled together.  One possibility would be to integrate
 out the  unphysical fluctuation first,  getting a (non-local) effective  Lagrangian for the
 physical  fluctuations. Here  we  shall treat all fluctuations on an equal footing; an
 alternative approach will be discussed in the next subsection. 
 
Labeling the bosonic fluctuations as 
\be
\Phi _I =\{ \Phi_{A\, i},\Phi_{S\, j} \} 
\ , \ \ \ \ \ \ \ \ 
\Phi_{A\, i}=\{ a_1,a_2,c \}  \,,~~~~ \Phi_{S\, i}=\{ b_1,b_2,d \} \,,
\ee
we find the following (euclidean) 
 bosonic propagator in the ``A'' and ``S'' sectors: 
\be \label{pads13}\ba{c}
 \triangle_{A}^{-1}(q) = 
\frac{1}{D_2} \left(
\begin{array}{ccc}
 -\frac{q^2}{2 } & -\frac{\hat{\kappa}^2 q_{_-} + i q^1 \hat \mu ^2}{ \hat \mu } 
 & -\frac{ \hat{\kappa} q_{_-}   \sqrt{\hat{\kappa}^2 -\hat \mu ^2}}{ \hat \mu 
  } \\
 \frac{\hat{\kappa}^2 q_{_-} + i q^1 \hat \mu ^2}{ \hat \mu  } & \frac{q_{_+} 
 \left(q_{_+}^2 \hat \mu ^2 - 4 \hat{\kappa}^2 \left(\hat{\kappa}^2- \hat
  \mu ^2\right)\right)}{2 q_{_-} \hat \mu ^2} & \frac{ 2 \hat \kappa  \sqrt{ 
  \hat{\kappa}^2 -\mu ^2} \left( \hat{\kappa}^2 q_{_-} +i q_1 \hat \mu ^2\right)}
  {q_{_+}  \hat \mu ^2} \\
 \frac{ \hat{\kappa} q_{_-}   \sqrt{\hat{\kappa}^2-\hat \mu ^2}}{ \hat \mu } &
  \frac{ 2 \hat{\kappa}  \sqrt{\hat{\kappa}^2 -\hat \mu ^2} \left(  \hat{\kappa}^2 
  q_{_-} +iq_1 \hat \mu ^2\right)}{q_{_+} \hat \mu ^2} & \frac{q^4 \hat \mu ^2-4 
  \left(q^1 \hat \mu ^2-i \hat{\kappa}^2 q_{_-} \right)^2}{2 q^2 \hat \mu ^2}
\end{array}
\right)  \,, \\[2pt]
\triangle_{S}^{-1} (q)=  {\rm diag}\left( -\frac{1}{2 \left(q^2+\hat\mu ^2\right)},
-\frac{1}{2 \left(q^2+\hat\mu ^2\right)},-\frac{1}{2 q^2} \right) \,,
~\\[3pt]
q_{_{\pm}}=q_0\pm i q_1\,,~~~~q^2=q_0^2 + q_1^2\,,~~~~D_2=q^4+4 \hat{\kappa}^2 q^2
 -4 \hat \mu ^2 q_1{}^2 \,,
\ea\ee
where  we rescaled 2d momentum $q$ by $\elk$ and 
\be \hat \kappa = { \kappa \ov \elk} \ , \ \ \ \ \ \ \ \ \ \ 
\hat \mu = { \mu \ov \elk} \ \la{fgf}
\ee
are the parameters that are fixed in the limit $\elk \to \infty$. 
The fermionic propagator following from \rf{ferpw3} 
has similar structure.  
The two-loop  graphs to be computed were described in section 
2.3. In the limit $\mu\to 0$ (i.e.  $\hat \mu\to 0$, $\hat \kappa \to 1$) 
we are interested in 
the physical  mass spectrum includes one bosonic mode with $m^2=4$  and
4 fermionic modes  with $m^2=1$ plus 3 massless bosonic modes.
 The $\mu\to 0$ limit  can be taken 
once  we simplify the integrands of  the two-loop integrals.
The general structure of the  two-loop integrals 
 appearing in the contributions of  the sunset diagrams 
in Figure \ref{feybos}(a) and Figure \ref{feyfer}(a) is  
\be \label{wmmm3}
	{\cal I}_{m_i^2,m_j^2,m^2_k}= \int \frac{d^2 \! q_i\, d^2\! q_j\, d^2\!
	 q_k}{(2\pi )^4}\frac{\mathcal{F}(q_i,q_j,q_k)}{q_i^{n_i}q_j^{n_j}q_k^{n_k}
	 (q^2_i+m_i^2)(q^2_j+m_j^2)(q^2_k+m_k^2)} \,.
\ee
while the double-bubble diagrams 
  in Figure \ref{feybos}(b) and Figure \ref{feyfer}(b)  lead to 
\be \label{wmm3}
	{\cal I}_{m_i^2, m_j^2}= \int \frac{d^2 \! q_i\, d^2\! q_j}{(2\pi )^4}\frac{\mathcal
	{F}(q_i,q_j)}{q_i^{n_i}q_j^{n_j} (q^2_i+m_i^2)(q^2_j+m_j^2)}\,.
\ee
Here $m^2_i,m^2_j,m^2_k$ can take values 0,1 or 4 and 
$\mathcal{F}$ are  some polynomial  functions of momenta.
The absence of modes with $m^2=2$  in the \ads case  suggests 
 the absence of contributions
proportional to the Catalan's constant.
Using an analytic regularization scheme  and 
   tensor manipulations  described in  \cite{grrtv}
   all  the integrals can be expressed  in terms of products
of two $I[m^2]$  factors  in  \eqref{uvandfin2}.
In addition to the 1PI diagrams there 
are non-1PI diagrams in  Figure \ref{feytad}. 
Since the bosonic propagator
  in \eqref{pads13} vanishes for $q=0$, one   is to define them 
   by first keeping  the  momentum of  the intermediate 
   bosonic line non-zero  and setting it to zero only after doing the one-loop 
   integrals.

  The resulting expression for the two-loop  effective action can be written as 
  \be\la{qw}
	\Gamma^{(2)} =\frac{8 \pi}{k_3}   V_2  \sum_n J_n \,,
\ee 
where   $J_n$    are contributions of different types of diagrams:
%
\be \label{pwa} \ba{l}
 J_{_{\rm boson~sunset}}=-\frac{1}{24} (6 I[4]I[0]+6 I[4]I[4])\,, \\ 
  J_{_{\rm boson~double-bubble}}=\frac{1}{16} (-4 I[4]I[0] -4 I[4]I[4])\,,
\\
  J_{_{\rm fermion-boson~sunset}}=\frac{1}{8} ( 12 I[1]I[0]+4 I[4]I[1]-8 I[1]I[1])\,,
  \\
   J_{_{\rm fermion-boson~double-bubble}} =-\frac{1}{8} (8 I[1]I[0]-8 I[4]I[1])\,, 
   \\
  J_{_{\rm boson-boson~tadpole}}=-\frac{1}{16} (-\frac{8}{3} I[4]I[0]-\frac{16}{3}
  I[4]I[4])\,,\\
     J_{_{\rm boson-fermion~tadpole}}=\frac{1}{16} (-\frac{8}{3} I[1]I[0]-\frac{40}{3} 
 I[4]I[1])\,,\\ 
  J_{_{\rm fermion-fermion~tadpole}}=-\frac{1}{16} (-8 I[1]I[1])\,.
 \ea\ee
 The sums  of the 1PI and non-1PI (tadpole)  contributions are given by
 %
 \bea \label{jus} &&
 J_{\rm 1PI}= -\frac{1}{2} \big(I[4] - I[1]\big)  \big(I[4] +  I[0] - 2 I[1]  \big)
 \ , \\ && 
 J_{\rm tadpole}= \frac{1}{6} \big(I[4] - I[1]\big)  \big(2I[4] + I[0] - 3 I[1]\big)
 \ .  \la{jas}\eea
 These  are separately UV finite  but IR divergent due to the presence of $I[0]$.  
 Notice also that both expressions are proportional to the coefficient 
 $I[4]-I[1]$ appearing in the one-loop  result \rf{jok}.
 
The total coefficient   is then (using \rf{mtoone3})
%
\bea && \sum_n J_n 
= -\frac{1}{6} \big( I[4] - I[1]\big)  \big(I[4] +2 I[0]-3 I[1]\big) \no
\\
&&\ \ \ \ \ \ \ \ \ = - \frac{1}{24 \pi^2} \left( \ln 2 \right)^2
	 -  \frac{1}{12 \pi ^2} \ln 2 \ln m_0   \ . \  \label{jj}  \eea
This expression 
is  still  IR divergent: we  introduced an IR cutoff $m_0\to 0$ 
 to define $I[0]$. 

While the presence of the  $\left( \ln 2 \right)^2$  contribution 
(absent in the corresponding string theory result found in Appendix \ref{ads3s3ost})
is an unambiguous result,   the appearance 
 of  the    IR divergence  should be  an artifact of our 
 computational procedure. It may be related, 
in particular, to  mixing between  massless  unphysical
and  physical  modes 
and/or to a possible ambiguity in how the limit $\mu\to 0$ was  taken. 
To support this expectation, in the next subsection 
 we shall repeat the above two-loop  computation  using a different  approach:
 by first integrating out $A_\pm$ and  gauge-fixing $g$ so that to explicitly
 eliminate all unphysical
 (non-coset) degrees of freedom  from the fluctuation Lagrangian.
The resulting two-loop correction will be found to be IR finite.


\subsection{Approach  based on integrating out gauge fields \\ and 
gauge-fixing  $g$ \label{ads3s3sg}}

In the case of the reduced theory for the $AdS_3 \times S^3$ string it is straightforward 
to integrate out $A_\pm$  and gauge-fix $g$  to get  an 
action  for the physical degrees  of freedom only, 
 which may then  be used for  computing the two-loop correction.
 The resulting  action 
is the  sum of the 
complex sinh-Gordon  and the complex sine-Gordon models  coupled to 
 two-dimensional fermions  \ci{gt2}.
 Depending on whether  one starts with the axial-gauged or vector-gauged WZW 
 model one gets the  ``tanh-tan'' (t-t) model or the ``coth-cot'' (c-c) model
 (these names refer to functions in the kinetic terms of the 2+2 coset bosonic 
 degrees of freedom). The two  models are related by the 2d duality and lead to equivalent 
 results for the partition  function when expanded near  the  respective 
  classical 
 solutions   corresponding to the long folded string 
 (i.e. near \rf{jh}  or its 2d dual analog). 
 It is useful to  consider  both  models in parallel  as this provides an extra check 
 on our computation. 
 The Lagrangians of the two models are 
 (the action is normalized as $I={ k \ov 8 \pi}  \int d^2 \sigma \ {\cal L}=
 { k_3 \ov 4 \pi}  \int d^2 \sigma \ {\cal L}$
 )
\be \label{lltanhtan} \ba{l}
	{\cal L}_{\rm t-t}=\partial_+ \varphi \partial_- \varphi + \tan ^2 \varphi\ \partial_+
	 \theta \partial_- \theta 
	+\partial_+ \phi \partial_- \phi + \tanh ^2 \phi\ \partial_+ \chi \partial_- \chi 
	+\frac{\mu^2}{2} \left( \cos 2 \varphi -\cosh 2 \phi \right) \\
	\hspace{10pt} +\alpha \partial_- \alpha +\beta \partial_- \beta +\gamma \partial_- 
	\gamma +\zeta \partial_- \zeta 
	+\lambda \partial_+ \lambda +\xi \partial_+ \xi +\rho \partial_+ \rho +\sigma \partial_+
	 \sigma \\
	\hspace{10pt} +\tan ^2 \varphi \left[ \partial_+ \theta \left( \lambda \xi -\rho \sigma
	 \right) - \partial_- \theta \left( \alpha \beta - \gamma \zeta \right) \right] 
	-\tanh ^2 \phi \left[ \partial_+ \chi \left( \lambda \xi -\rho \sigma \right) - \partial_-
	 \chi \left( \alpha \beta - \gamma \zeta \right) \right] \\
	\hspace{10pt} -\left( \alpha \beta - \gamma \zeta \right) \left( \lambda \xi -\rho \sigma 
	\right) \left( \frac{1}{\cos ^2 \varphi}-\frac{1}{\cosh ^2 \phi} \right)
	-2 \mu \Big( 
	\cosh \phi \cos \varphi \left( \lambda \gamma + \xi \zeta -\rho \alpha -\sigma \beta \right) \\
	\hspace{10pt} + \sinh \phi \sin \varphi \big[ \cos ( \chi + \theta) \left( -\rho
	 \zeta + \sigma \gamma + \lambda \beta -\xi \alpha \right)
	 \\
	\hspace{10pt}
	  - \sin ( \chi + \theta) \left(
	  \lambda \alpha + \xi \beta+ \rho \gamma + \sigma \zeta \right) \big] \Big) \,,
\ea \ee 
\be \label{llcothcot} \ba{l}
	{\cal L}_{\rm c-c}=\partial_+ \varphi \partial_- \varphi + \cot ^2 \varphi\ \partial_+ 
	\theta \partial_- \theta 
	+\partial_+ \phi \partial_- \phi + \coth ^2 \phi\ \partial_+ \chi \partial_- \chi 
	+\frac{\mu^2}{2} \left( \cos 2 \varphi -\cosh 2 \phi \right) \\
	\hspace{10pt} +\alpha \partial_- \alpha +\beta \partial_- \beta +\gamma \partial_- \gamma 
	+\zeta \partial_- \zeta 
	+\lambda \partial_+ \lambda +\xi \partial_+ \xi +\rho \partial_+ \rho +\sigma \partial_+ 
	\sigma \\
	\hspace{10pt} -\cot ^2 \varphi \left[ \partial_+ \theta \left( \lambda \xi -\rho \sigma
	 \right) - \partial_- \theta \left( \alpha \beta - \gamma \zeta \right) \right] 
	+\coth ^2 \phi \left[ \partial_+ \chi \left( \lambda \xi -\rho \sigma \right) -
	\partial_- \chi \left( \alpha \beta - \gamma \zeta \right) \right] \\
	\hspace{10pt} -\left( \alpha \beta - \gamma \zeta \right) \left( \lambda \xi -\rho \sigma
	 \right) \left( \frac{1}{\sin ^2 \varphi}+\frac{1}{\sinh ^2 \phi} \right) -2 \mu \Big( 
	 \sinh \phi \sin \varphi \left( \lambda \gamma + \xi \zeta -\rho \alpha -\sigma \beta \right) 
	 \\
	\hspace{10pt} + \cosh \phi \cos \varphi \big[ \cos ( \chi + \theta) \left( \rho \zeta
	 - \sigma \gamma - \lambda \beta +\xi \alpha \right) \\
	 \hspace{10pt}- \sin ( \chi + \theta) \left( \lambda 
	 \alpha + \xi \beta+ \rho \gamma + \sigma \zeta \right) \big] \Big) \,.  
\ea \ee 
Here  $\phi,\,\theta$  correspond to  bosonic degrees of freedom related to $AdS_3$, 
 $\varphi,\,\chi$  correspond to $S^3$ part  and $\alpha,\,\beta,\,\gamma,\,
  \zeta,\,\lambda,\,\xi,\,\rho,\,\sigma$ are real fermionic fields.  

For technical reasons (to make the expansion near the  vacuum point regular)
it is useful to generalize  the  reduced theory solution 
\rf{jh} 
 by introducing also a similar non-trivial  background in the ``$S^3$'' part of the model. 
 Namely,  we may start  with the reduced theory background   corresponding to 
 the  following generalization of the long spinning string in \rf{sj}:
\be\la{sjfoldsc}
\ba{c}
	Y_0+iY_{5}=\cosh (\elk \sigma) \,e^{i\kappa \tau}\,, ~~~~~~ Y_1+iY_2= \sinh
	 (\elk \sigma) \,e^{i\kappa \tau}\,,\ \ \ \ \ \ \   Y_{3,4}=0 \ ,  \\	
	X_1+iX_{2}=\frac{1}{\sqrt{2}} e^{i \omega \tau + i {\rm n} \sigma}\,, ~~~~~~X_3
	+iX_{4}=\frac{1}{\sqrt{2}} e^{i \omega \tau - i {\rm n} \sigma}\,, \ \ \ \ 
	X_{5,6}=0 \ , \\ 
	\kappa^2 = \elk^2+ \mu^2 \,,~~~~~~~~~~~~~~~\mu^2 = {\rm n}^2 + \omega^2 \,.
\ea\ee
 This   solution represents a superposition 
 of a string with large spin in $AdS_3$ and a circular string  with 
 two large equal spins in $S^3$. 
 The corresponding  classical solutions in tanh-tan  and coth-cot 
 models (related again by 2d duality)  are \ci{iwa}
\be\label{solredtan}\ba{l} {\rm{t-t}}: \ \ \ \ 
	\phi_0 = \ln \frac{\kappa +\sqrt{\kappa ^2-\mu ^2}}{\mu } \,,~~~\ \ \ \ \ \ \ \ \ \ \ \ ~~\chi_0
	 =\frac{\kappa ^2}{\mu } \tau \,,\\ \ \ \ \ \  \ \ \ \ \ \ \ \ 
	\varphi_0= \frac{1}{2} \text{arccos}\left( \frac{2 \omega ^2}{\mu ^2}-1 \right) \,,
	~~~\ \ \ \ ~~\theta_0 = \frac{\omega ^2}{\mu } \tau \,,
\ea\ee
\be\label{solredcot}\ba{l} \ \ {\rm{c-c}}: \ \ \ \ 
	\phi_0 = \ln \frac{\kappa +\sqrt{\kappa ^2-\mu ^2}}{\mu } \,,~~\ \ \ \ \  \ \ \ \ \ \ ~~~\chi_0 
	=\frac{\mu^2-\kappa ^2}{\mu } \sigma \,,\\ \ \ \ \ \ \ \ \ \ \ \ \ \ \ \ 
	\varphi_0= \frac{1}{2} \text{arccos}\left( \frac{2 \omega ^2}{\mu ^2}-1 \right) \,,~~\ \ \ \ ~~~
	\theta_0 = \frac{\omega ^2-\mu^2}{\mu } \sigma \,.
\ea\ee
Below  we shall consider the  one-loop and two-loop corrections in the models
\rf{lltanhtan} and \rf{llcothcot} expanded  near these solutions. 
We will  eventually be  interested in the limit 
 \be\la{kp}
  \mu\to 0,\ \ \ \ \ \ \ \  \o	\to 0,\ \ \ \  \ \ \  \ \k \to \elk \ \gg 1  \ , \ \ \  \   \ \ \ 
   \ {\rm n}\to 0 \ .  \ee

\subsubsection{One-loop contribution}

Expanding  the bosonic parts of the reduced theory Lagrangians 
\eqref{lltanhtan} and  \eqref{llcothcot} 
near the classical solutions \eqref{solredtan}  and  \eqref{solredcot},
\be\ba{c}
	\phi=\phi_0+\delta \phi \,,~~~~~\chi= \chi_0+ \delta \chi \,,\ \ \ \ \ \ \ \ \ 
	\varphi=\varphi_0+\delta \phi \,,~~~~~\theta= \theta_0+ \delta \theta  \,,
\ea\ee
 leads to the following quadratic fluctuation Lagrangians 
\be\ba{l}
\displaystyle {\cal L}_{\rm t-t}^{(2B)}= \partial_- \delta \phi \partial_+ \delta \phi+\partial_- 
\delta \varphi \partial_+ \delta \varphi - 4 \left(\kappa ^2-\mu ^2\right) \delta \phi^2-4 \left(\omega 
^2- \mu ^2\right) \delta \varphi^2 
\displaystyle + \frac{\kappa ^2-\mu ^2}{\kappa ^2} \partial_- \delta \chi \partial_+ \delta \chi \\ 
 \hspace{10pt}
\displaystyle + \frac{2 \mu  \sqrt{\kappa ^2-\mu ^2} }{\kappa }  \delta \phi \left(\partial_- \delta 
\chi+\partial_+ \delta \chi\right)
+ \frac{\mu ^2-\omega ^2 }{\omega ^2} \partial_- \delta \theta \partial_+ \delta \theta +\frac{2 \mu 
 \sqrt{\mu ^2-\omega ^2} }{\omega }  \delta \varphi \left(\partial_- \delta \theta+\partial_+ \delta 
 \theta\right) \,,
\ea\ee
\be\ba{l}
\displaystyle {\cal L}_{\rm c-c}^{(2B)}= \partial_- \delta \phi \partial_+ \delta \phi+\partial_- \delta \varphi 
\partial_+ \delta \varphi - 4 \kappa ^2 \delta \phi^2-4 \omega ^2 \delta \varphi^2 
\displaystyle +\frac{\kappa ^2 }{\kappa ^2-\mu ^2} \partial_- \delta \chi \partial_+ \delta \chi \\  
\hspace{10pt} 
\displaystyle +\frac{2 \kappa  \mu  }{\sqrt{\kappa ^2-\mu ^2}}  \delta \phi \left(\partial_- \delta
 \chi-\partial_+ \delta \chi\right)
+\frac{\omega ^2 }{\mu ^2-\omega ^2} \partial_- \delta \theta \partial_+ \delta \theta +\frac{2 \mu 
 \omega  }{\sqrt{\mu ^2-\omega ^2}}  \delta \varphi \left(\partial_- \delta \theta-\partial_+ \delta 
 \theta\right) \,.
\ea\ee 
The resulting  bosonic factors in the one-loop partition functions 
are equal and are also  the same as  in the original string theory\foot{The corresponding 
 characteristic frequencies of the $4$ bosonic fluctuations  are  
$	\sqrt{n^2+2 \kappa ^2\pm 2 \sqrt{\kappa ^4+n^2 \mu ^2}} $ and $ 
	\sqrt{n^2+2 \omega ^2 \pm 2 \sqrt{n^2 \mu ^2+\omega ^4}}$.
	These  match the string-theory expressions  \cite{ftt,art}.}  
\bea 
&& Z^{(1 B)}_{ \rm t-t}= Z^{(1 B)}_{ \rm c-c} = \Big(\det 
 \left[ \partial_+ ^2 \partial_- ^2 + 2(2 \kappa^2 -\mu^2) \partial_+ \partial_- +\mu^2
  (\partial_+ ^2 +  \partial_- ^2) \right]\Big)^{-1/2} \no \\
  &&\ \ \ \  \ \ \ \ \ \ \ \ \ \ \ \ \  \ \ \ \times \Big( \left[( 
  \partial_+ ^2 \partial_- ^2 + 2(2 \omega^2
   -\mu^2) \partial_+ \partial_- +\mu^2( \partial_+ ^2 +  \partial_- ^2) \right] \Big)^{-1/2}\,,
\eea
To  simplify the fermionic Lagrangians in  \eqref{lltanhtan},
 \eqref{llcothcot} (making the coefficients in them constant) it is useful to 
 rotate the fermionic fluctuations in the following way 
\bea && 
 \alpha + i \beta \to (\alpha + i \beta) e^{\cal B} \,,\ \ 
 \gamma + i \zeta \to (\gamma + i \zeta ) e^{{\cal B}^*} \,,\ \ 
 \lambda + i \xi \to (\lambda + i \xi ) e^{{\cal B}^*} \,,\ \ 
 \rho + i \sigma \to (\rho + i \sigma ) e^{\cal B} \,, \no \\
 && {\cal B}_{\rm t-t}
 =i \frac{ \kappa ^2+\omega ^2 }{2 \mu } \tau \ , \ \ \ \ \ \ \ \ \ \ \ \ 
 {\cal B}_{\rm c-c}=i \frac{ \kappa ^2-\omega ^2}{2 \mu } \sigma  \ . \label{rotfer2}
 \eea
 The corresponding fermionic one-loop determinants are then found to be 
\bea &&
Z^{(1 F)} _{\rm t-t}=\Big[\det \big[ \partial_+^2 \partial_-^2+
\mu ^2\left(\partial_+^2 +\partial_-^2\right)
  +2 \left(\kappa ^2-\mu ^2+\omega ^2\right)\partial_+ \partial_- +\left(\kappa ^2-\omega 
  ^2\right)^2\big]\Big]^2\no \label{detfermtt}\\ 
&&
 Z^{(1 F)} _{\rm c-c}=\Big[\det \left( \partial_+ \partial_-
  +\kappa ^2-\mu ^2+\omega ^2\right)\Big]^4\,. 
\eea
Despite looking different, these two expressions can be shown to be  equivalent.

In the limit $\mu,\o\to 0$ we recover the expression found in the approach of section 3.1 
equal also to the string theory result: 
\be 
Z^{(1)}= Z^{(1B)}Z^{(1F)}=
 \big[\det( \partial_+ \partial_- +4 \kappa^2 )\big]^{-1/2} 
  \big[\det( \partial_+ \partial_-)\big]^{-3/2} 
\big[\det( \partial_+ \partial_- +  \kappa^2 )\big]^{4} 
\ . \ee

\subsubsection{Two-loop contribution}

\begin{figure}[tb]
 \begin{center}
  \includegraphics[height=3.cm]{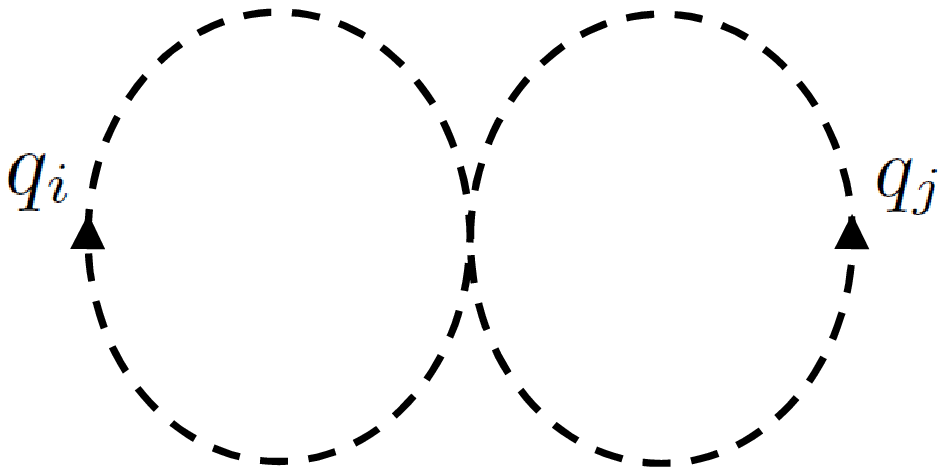}
 \end{center}
 \caption{Fermionic double-bubble diagram.}
 \label{feyadd}
\end{figure}

Since  in the present case the   Lagrangians \eqref{lltanhtan}, \eqref{llcothcot} 
found after integrating out gauge fields in \rf{dwzwac} 
contain quartic fermionic terms,  
in addition to diagrams in  Figure \ref{feybos} 
 and Figure \ref{feyfer} we will also have to compute 
 the  fermionic double-bubble diagram in 
Figure \ref{feyadd}.\foot{One may wonder also if we should account for a local one-loop
counterterm  \ci{hts}  originating from integrating out $A_\pm$.  This counterterm 
 leads, however,
 only to  power-divergent two-loop corrections which (along with similar contributions from
 other diagrams)  are to be  regularized away.} 
Applying the redefinitions in   \eqref{rotfer2} one finds  that 
the   coefficients in the  cubic and quartic fermionic terms in the fluctuation Lagrangian
 are  constant.  Hence 
the  computation of two-loop corrections 
is similar to that  in the first approach in section 3.1 

 As we are interested in the result in the 
 $\mu,\omega \to 0$ limit, one   should be careful to keep track of possible ambiguities 
 in taking this limit  that may be present in the individual diagrams by 
  introducing  the  parameter 
 \be r\equiv  {\omega\ov  \mu }  \ . \ee 
The results for the contributions of individual diagrams to $\Gamma^{(2)}$ in \rf{qw} 
are found to be (cf. \rf{pwa})
%
\be\label{j2sg1}\ba{l}
{\rm t-t}: \ \ \   J_{_{\rm boson~double-bubble}} = \frac{1}{16} \left( -8 I[4] I[4]\right)\,,\\
\ \ \ \  \ \ \ \ \ \ \  \ J_{_{\rm fermion-boson~sunset}} = \frac{1}{8} \left( 8  I[1] 
I[0] +  4 I[4] I[1] -
 2 \frac{1 + 2 r^2}{r^2} I[1] I[1]\right)\,,\\
\  \ \ \ \ \ \ \  \ \ \ \ J_{_{\rm fermion-boson~double-bubble}} = -\frac{1}{8} \left( 
8  I[1] I[0] - 4  I[4]
 I[1]\right)\,,\\ \ \ 
\ \  \ \ \ \ \  \ \ \ J_{_{\rm fermion-fermion~double-bubble}} = \frac{1}{16} \left(
 4 \frac{1 - 2 r^2}{r^2} 
I[1] I[1]\right)\,,\\ \ \ \ \ \ \ \ \ \ \ \  \ 
J_{_{\rm tadpole}} = -\frac{1}{16} \left( -8 I[1] I[1]\right)\,,
\ea\ee
%
%
\be\label{j2sg2}\ba{l}
{\rm c-c}: \ \ \ J_{_{\rm boson~double-bubble}}=  \frac{1}{16} \left( -8  I[4] I[4]\right)\,,\\
\ \ \ \ \ \ \ \   \ \ \ \ 
J_{_{\rm fermion-boson~sunset}} = \frac{1}{8} \left( 8  I[1] I[0] +  4 I[4] I[1] - 
\frac{6-4r^2}{1-r^2} I[1] I[1]\right)\,,\\ \ \  \ \ \ \ \  \ \ \ \ \ 
J_{_{\rm fermion-boson~double-bubble}} = -\frac{1}{8} \left( 8 I[1] I[0] - 4  I[4]
 I[1]\right)\,,\\ \ \ \ \ \  \ \ \ \ \  \ \ 
J_{_{\rm fermion-fermion~double-bubble}} = \frac{1}{16} \left( -4 \frac{1-2 r^2}{1-r^2}
 I[1] I[1]\right)\,,\\ \ \ \ \ \  \ \ \ \ \   \  \ 
J_{_{\rm tadpole} } = -\frac{1}{16} \left(-8 I[1] I[1]\right)\,.
\ea\ee
Summing up the 1PI  contributions   we   find that IR-divergent and $r$-dependent
 terms cancel out and we
get the same result in the two models:
 %
\bea \label{kus}
 && J_{\rm 1PI}=  - \frac{1}{2}I[4] I[4]  +  I[4] I[1] -  I[1] I[1] \ , \\
 && J_{\rm tadpole}=  \frac{1}{2}I[1] I[1] 
 \ , \label{kas} \eea
so that the total is (using \rf{mtoone3})
\be \la{kpq}
\Gamma^{(2)} =\frac{8 \pi}{k_3}   V_2  \sum_n J_n \,, \ \ \ \ \  \ \ \ \ \ \ \ \ 
\sum_n J_n = - \frac{1}{2}\big(I[4]-I[1]\big)^2 = -\frac{1}{8\pi^2 }\left(  \ln 2 \right)^2  \ . 
\ee
%
Combining everything together 
we find that the effective action for the \ads model is (cf. \rf{ads})
%
\be 
\Gamma^{(2)} = \frac{1}{\pi k_3}\ \aa_2 \ V_2 \ , \ \ \ \ \ \  \ \ \ \ \ \ 
\aa_2 = - {1 \ov 4} (\aa_1)^2 = - (\ln 2)^2  \  . 
\ee
Let us  now compare these results with those  \rf{jus},\rf{jas},\rf{jj} 
found in the first approach in section 3.1.
We observe that 1PI  contributions in \rf{jus} and \rf{kus}
contain the same $I[4] I[4]$ and $I[1] I[1]$ terms;\foot{Note that in contrast to 
the result \rf{pwa} in the  first approach in section 3.1 
here we have no  bosonic sunset contribution
but the bosonic double-bubble 
contribution gives the same $I[4]I[4]$ term as the sum of the bosonic sunset 
and double-bubble contributions in \rf{pwa}.
 This should not be too  surprising 
as the contributions of individual diagrams may be different in different gauges.}
 also,  the  fermion-fermion $I[1] I[1]$ tadpole  terms are the same. 
  Given  that 
 the final result should be UV and IR finite,   then the expression in 
 \rf{kpq} is a  natural outcome.  This   suggests 
 that it is the tadpole  contribution \rf{jas} in the first  approach 
 that is to be  blamed  for  the IR  problem found there: it should not actually  contain 
the  $I[4] I[4]$ term if the two  approaches are to agree. 
 Then if instead of \rf{jus} one would  take 
 %
\be
 \J'_{\rm tadpole}= \frac{1}{6} \big(I[4] - I[1]\big)  \big( 3 I[0]-3 I[1]\big)
=  \frac{1}{2}\big(I[4] - I[1]\big)  \big( I[0]- I[1]\big)
 \ ,  \la{jass}\ee
then the sum  of \rf{jass} with the 1PI contribution \rf{jus} in the first approach 
 would  exactly  match the
result \rf{kpq} of the second approach.\foot{The replacement  of 
\rf{jas} by \rf{jass} is formally achieved 
by  replacing $I[4]$ in the second factor in \rf{jas} by $I[0]$.
This may be related to a subtlety in   how  the two  massive  $AdS_3$ modes (which are 
mixed for $\mu \not=0$)
are treated in the tadpole contributions in the limit when $\mu\to 0$: 
 in that limit one of them has $m^2=4$ and the other one becomes  massless.}

It is also interesting to note    that the final two-loop  result in \rf{kpq}
is  proportional to the square   of the coefficient  in the 
 one-loop  contribution in \rf{jok}.
  These observations will guide us in interpreting and fixing  the two-loop 
 result in the case of the reduced theory for \adss  string  
 where we will only  have the expression  found using   the first approach.


\section{Reduced theory for ${ AdS}_5 \times { S}^5$ string\label{secredads5}}

Let us now carry out the similar  computation in the reduced ${ AdS}_5 \times { S}^5$ theory. 
Here  there are  more fields and following the second 
approach based on integrating 
out  gauge fields first appears to be difficult. 
For that  reason here  we will follow the first approach 
described in section 2.2. We will unambiguously determine the coefficient of the 
Catalan's constant term  and match it with the string theory result.
As in the ${ AdS}_3 \times { S}^3$ 
case discussed in section 3.1, in 
this approach there will be a non-canceling IR  divergence 
which should be an artifact of mixing of physical and unphysical   modes 
in this approach. 
A close analogy  
 with the ${ AdS}_3 \times { S}^3$  case will  motivate 
 a modification of the tadpole contribution that will
lead to  IR finite  two-loop  $(\ln 2)^2$ term.

The  reduced theory solution   corresponding to the long 
 $(S,J)$ folded string  \eqref{sj}  here has a similar structure to \rf{gag},\rf{hag}.
 Following  
\cite{hit,iwa},  here we shall  choose it in the $\tau$-dependent 
 form (cf. \rf{gag})\foot{
In general, a  
choice of the reduced theory  solution 
corresponding to a given string theory solution
 is not  unique as one may apply an 
on-shell $H\times H$ gauge transformation. 
For example, one may  start with a $\sigma$-dependent solution, 
\by \nonumber
g_A^\prime =\left(
\begin{array}{cccc}
 \frac{ \kappa }{\mu }v_{\sigma}^* & 0 & 0 & \frac{\ell }{\mu } \\
 0 & \frac{ \kappa }{\mu } v_{\sigma} & \frac{\ell }{\mu }  & 0 \\
 0 & \frac{\ell }{\mu } & \frac{ \kappa }{\mu } v_{\sigma}^* & 0 \\
 \frac{\ell }{\mu } & 0 & 0 & \frac{ \kappa }{\mu } v_{\sigma}
\end{array}
\right) \,,~~~~~~~~~~~~
	A_{+ A}^\prime = A_{- A}^\prime =\frac{i \kappa ^2}{2 \mu } \Sigma \,.
\ey
One may  expect
 that the result for the quantum partition function
 for the two solutions should be the same. 
In fact,
we have checked  that 
 the individual diagram contributions in the two cases are indeed  the same. 
}
\bea\label{clagane0} && 
g_0=\Big( \ba{cc}
g_A&{\bf 0}\\ {\bf 0}&{\bf 1} \ea
\Big)\,, ~~~~~
g_A=\left(\ba{cccc}
	 0 & \frac{\kappa}{\mu} v^*_{\tau} & -\frac{\elk}{\mu} v^*_{\tau}  & 0
	\\  -\frac{\kappa}{\mu} v_{\tau}  &0 &0 &  \frac{\elk}{\mu} v_{\tau}
	\\  \frac{\elk}{\mu} v^*_{\tau}  &0 &0 &  -\frac{\kappa}{\mu} v^* _{\tau}
	\\ 0& -\frac{\elk}{\mu} v_{\tau} & \frac{\kappa}{\mu} v_{\tau} & 0
	\ea\right) ,~~~~~~~~
v_{\tau}=  e^{i\frac{\kappa^2\tau}{\mu}}  \\
&&	\label{claane0}
 A_{\pm 0}=\Big( \ba{cc}
A_{\pm A}&  {\bf 0}  \\ {\bf 0} & {\bf 0}  \ea
\Big)\, ,~~~~
	A_{+A}=\frac{i(\elk^2+\kappa^2)}{2\mu}\ {\Sigma} \ , \ \ \ 
	A_{-A}=\frac{i\mu}{2}{\Sigma} \ , \ \ \ 
	\Sigma = {\rm diag}(1,-1,1,-1)  \ .
\eea
The corresponding  solutions  for 
	the fields  $\td g, \ U, \td U$ in \eqref{claslag},\rf{atou},\rf{ree}
	 are (cf.
	\rf{utu3},\rf{g3pw})
\bea &&
U_0=\Big(\ba{cc}
u&0\\
0&{\bf 1}
\ea \Big)
  \,, ~~~~~~~~~~~~~~~
\tilde{U}_0= \Big(\ba{cc}
\tilde{u}&0\\
0&{\bf 1} 
\ea \Big) \,,\\  &&
u=\tilde u=\left(
\begin{array}{cccc}
 w^{*1/2} & 0 & 0 & 0 \\
 0 & w^{1/2} & 0 & 0 \\
 0 & 0 & w^{*1/2} & 0 \\
 0 & 0 & 0 & w^{1/2}
\end{array}
\right) \,, \ \ \ \ \ \
	w\equiv v_\tau v_\sigma =e^{i \frac{ \kappa ^2 \tau + \ell^2 \sigma }{\mu }} \,,
\\
&&
\tilde{g}_0= U^{-1}_0 g_0 \td U_0 = \Big(\ba{cc}
 \tilde{g}_A & 0 \\
 0 & 1
  \ea
  \Big) \,,\ \ \ \ \ \ \ \ 
\tilde{g}_A=
\left(
\begin{array}{cccc}
 0 & \frac{ \kappa }{\mu } v_\sigma & -\frac{ \elk }{\mu } v_\tau ^* & 0 \\
 -\frac{ \kappa }{\mu } v_\sigma ^* & 0 & 0 & \frac{ \elk }{\mu } v_\tau \\
 \frac{ \elk}{\mu } v_\tau^* & 0 & 0 & -\frac{ \kappa }{\mu } v_\sigma \\
 0 & -\frac{ \elk}{\mu } v_\tau & \frac{ \kappa }{\mu } v_\sigma^* & 0
\end{array}
\right) \,.\la{gog}
\eea
Since $U_0^{-1} \tilde{U}_0= 1$, the two-loop contribution of the WZW term
 $I[U^{-1} \tilde{U}]$
in \rf{claslag}  will be trivial. 
The value of the classical reduced theory action 
 on this background is  ($k=k_5$) 
\be
\Gamma^{(0)}=I_\pr= \frac{k}{4\pi}V_2\,\frac{(\kappa^2-\mu^2)^2}{\mu^2} \ .
\ee

 \iff
and, by computing the stress tensor for the solution \eqref{sj}, one finds that 
the mass scale in the reduced theory is $\mu$.
Note that this parameterization of the classical solution corresponds to the
 axial gauging whereas we used the vector gauging in section \ref{ads3s3gwzw}.
  Consequently, the classical solution \eqref{clagane0} has the $\tau$ dependence. 
These two ways of the gauging are related by the $H\times H$ gauge transformation 
as shown in \cite{iwa}, the final result should be independent of the choice of the gauging.  
\fi

\subsection{One-loop contribution\label{seconeloop}}

The analysis of  quadratic fluctuation Lagrangian is similar to the one in section 3.1.1
though more involved as now the  fluctuation  fields  have more components. 
The resulting  one-loop partition function will match again the 
corresponding string theory result. 
\iff
The argument given here is based on the approach using the PW identity. As an
 alternative approach, the one-loop computation in the $\delta A_+=0$ gauge is 
 discussed in Appendix \ref{appda0}. Both the approaches give the same functional 
 determinant contribution, so they are consistent. 
\fi
To ensure that the   coefficients in the fluctuation Lagrangian \rf{pw2}
are constant we parametrize the fluctuation fields as follows:
\be\label{etapr}\ba{c}
	\eta^\parallel =\left(
	\ba{cc}
	\eta^\parallel _A & 0 \\
	0& \eta^\parallel _S
	\ea \right) \,,  \ \ \ \ \ 
	\eta^\parallel _A =\left(
	\begin{array}{cccc}
		0 & 0 & a_{1}+ia_{2}& (a_{3}+ia_{4}) w \\
		0 & 0 & (a_{3}-ia_{4}) w ^* & -a_{1}+ia_{2}\\
		a_{1}-ia_{2}& (a_{3}+ia_{4}) w & 0 & 0 \\
		(a_{3}-ia_{4}) w^* & -a_{1}-ia_{2}& 0 & 0  
	\end{array}
	\right)\,, \\
	\eta^\parallel _S=\left(
\begin{array}{cccc}
 0 & 0 & b_1+i b_2 & b_3+i b_4 \\
 0 & 0 & -b_3+i b_4 & b_1-i b_2 \\
 -b_1+i b_2 & b_3+i b_4 & 0 & 0 \\
 -b_3+i b_4 & -b_1-i b_2 & 0 & 0
\end{array}
\right) , 
\\
	\eta^\perp =\left(
	\ba{cc}
	\eta^\perp _A & 0 \\
	0& \eta^\perp _S
	\ea \right) \,,\ \ \ \ \ 
\eta^\perp _A=\left(\ba{cccc}
i c_1 & (c_2+i c_3) w&0&0
\\(-c_2+i c_3) w^* &-i c_1&0&0
\\0&0&i c_4 & ( c_5+i c_6 )w 
\\0&0&(-c_5+i c_6) w^* &-i c_4\ea\right)\, , \\
\eta^\perp _S=\left(
\begin{array}{cccc}
 i d_1 & d_2+i d_3 & 0 & 0 \\
 -d_2+i d_3 & -i d_1 & 0 & 0 \\
 0 & 0 & i d_4 & d_5+i d_6 \\
 0 & 0 & -d_5+i d_6 & -i d_4
\end{array}
\right) \,.
\ea
\ee
Here $\eta^\parallel$ represent  4+4  physical (coset) fluctuations and $\eta^\perp$ 
represent 6+6  unphysical fields.
The ``$S^5$'' part  of the fluctuation Lagrangian is simply 
\be  \label{s5total}
	{\cal L}_{{\rm S}}^{(2)}=2 \sum_{i=1}^4 \left(  \partial _+ b_i \partial _- 
	b_i - \mu^2 b_i^2 \right) + \sum_{j=1}^6 \partial _+ d_j \partial _- d_j \,,
\ee
describing 4 mass $\mu$ degrees of freedom as in string theory.  
The ``$AdS_5$'' part splits into the  subsectors of mixed fluctuations: 
one subsector
 contains $a_1$, $a_2$ and the off-diagonal part of $\eta^\perp _A$ --- $c_2$,
  $c_3$,$c_5$ and $c_6$; 
   the other  contains $a_3$, $a_4$ and the 
  diagonal part of $\eta^\perp _A$ --- $c_1$ and $c_4$. 
  Explicitly, 
\bea \label{adstotal}
	&&{\cal L}^{(2)}_{{\rm A}}={\cal L}^{(2)}_{1}+{\cal L}^{(2)}_{2}\,,\\
&&\label{la1a2}
{\cal L}^{(2)}_{1} =  2 \sum _{i=1,2} \left[  \partial_+ a_i \partial_- a_i 
- \left(2 \kappa ^2-\mu ^2\right) a_i^2  \right]  - 4 M_1 \left( \mu c_2 +  \partial_- c_3
 + \mu c_5+  \partial_-c_6 \right) a_1 \nonumber \\ 
&&\hspace{20pt} + 4 M_1 \left(  \partial_- c_2 -\mu c_3 -  \partial_- c_5 + 
\mu c_6 \right) a_2  
- \sum _{j=2,3,5,6} \left[  \partial_+ c_j \partial_- c_j +\left(2 \kappa ^2-\mu ^2\right) 
c_j^2 \right] \nonumber \\ 
&&\hspace{20pt}
-2 \left( \mu \partial_+ c_3 + M_2 \partial_- c_3  \right) c_2 
-2 \left( \mu \partial_+ c_6 + M_2 \partial_- c_6  \right) c_5
\ , \no \\
&&
  {\cal L}^{(2)}_{2} =  2 \sum _{i=3,4}  \partial_+ a_i \partial_- a_i + 4 \left( \mu 
  \partial_+a_4 + M_2 \partial_-a_4 \right) a_3  - \sum _{j=1,4}  \partial_+ c_j \partial_-
   c_j + 4 M_1 (\partial_- c_1+\partial_- c_4) a_3  \ , \no 
\eea
where the constants $M_1=\frac{\kappa \sqrt{\kappa^2 -\mu^2}}{\mu^2}$, $M_2=
\frac{  2\kappa^2-\mu^2}{\mu }$   are the same as  in \rf{bbdef3}.
The  bosonic contribution to the  one-loop partition function is then 
\bea &&
Z^{(1B)} = Z^{(1B)}_{\rm A}  Z^{(1B)}_{\rm S}  \ , \ \ \ \ \ 
Z^{(1B)}_{\rm S}
= \Big(   [\det
\left(\partial_+ \partial_-+\mu ^2\right)]^4 
\ [ \det \left( \partial_+ \partial_-\right)] ^6 \Big) ^{-1/2} \ , \\
&&
 \label{detads} 
Z^{(1B)}_{\rm A}	= \Big( 
[ \det \left(\partial_+ \partial_- +2 \kappa ^2-\mu ^2\right)]^{2} 
\ [\det (\partial_+ \partial_-)]^2\  [\det\left( \partial_-^2+\mu
	   ^2\right)]^2\  [\det \left( \partial_+^2+\mu ^2\right)]^2 \no
\\
&& \ \ \ \ \ \ \ \ \ \ \times  \det\left[ 
	 \partial_+^2 \partial_-^2 + 2 \left(2 \kappa ^2- \mu ^2\right) \partial_+
	  \partial_- +  \mu ^2( \partial_+^2 + \partial_-^2 )\right] \Big)^{-1/2} 
	  \ . \eea
Observing  that 
 \be 
 \det ( \partial_\pm^2+\mu ^2)  = \det ( \partial_\pm + i \mu ) \ \det 
	 ( \partial_\pm - i \mu ) 
 = \det ( e^{-i \mu \tau} \partial_\pm  e^{i \mu \tau} ) \det
	 ( e^{i \mu \tau} \partial_\pm  e^{-i \mu \tau} ) \,,
\ee
and accounting for  the massless determinants coming from the  Jacobian of 
 transformation \eqref{atou} and the quantum fluctuations 
 of $U$ in the    WZW term in \eqref{claslag}
 one  concludes that the bosonic part of the one-loop partition function is the same as 
 in the corresponding \adss string theory.
 
 The parametrization of the fermionic fluctuations 
\bea &&
	\Psi_{_R}=\left(
	\begin{array}{cc}
		0& \mathfrak{X}_R \\
		\mathfrak{Y}_R& 0 
	\end{array}
	\right) \,, ~~~~~~~~~~~~
	\Psi_{_L}=\left(
	\begin{array}{cc}
		0& \mathfrak{X}_L  \\
		\mathfrak{Y}_L& 0 
	\end{array}
	\right) \,, \no \\
	&&  \label{fxr}
	\mathfrak{X}_R=\left(
	\begin{array}{cccc}
		 0 & 0 &(\alpha_1+i\alpha_2) t_{_+} &(\alpha_3+i\alpha_4) t_{_+} \\
		 0 & 0 &(-\alpha_3+i\alpha_4) t_{_+}^* &(\alpha_1-i\alpha_2) t_{_+}^* \\
		(\alpha_5+i\alpha_6) t_{_+} &(\alpha_7-i\alpha_8) t_{_+} &0&0\\
		(\alpha_7+i\alpha_8) t_{_+}^* &(-\alpha_5+i\alpha_6) t_{_+}^* &0 &0 \\
	\end{array}
	\right)\,, \ \ \ \ \ \  t_{_{\pm}} \equiv  e^{i \frac{\kappa
 ^2 \tau \pm \elk ^2 \sigma }{2 \mu } }\label{fyl}\no   \\
&&	\mathfrak{Y}_R=\left(
	\begin{array}{cccc}
		0 & 0 &(-\alpha_6 -i\alpha_5) t_{_+}^* & (-\alpha_8 -i\alpha_7) t_{_+} \\
		0 & 0 &(\alpha_8 -i\alpha_7) t_{_+}^* & (-\alpha_6 +i\alpha_5) t_{_+} \\
		(\alpha_2 +i\alpha_1) t_{_+}^* & (\alpha_4 -i\alpha_3 ) t_{_+} &0 &0  \\
		(\alpha_4 +i\alpha_3) t_{_+}^* & (-\alpha_2 +i\alpha_1) t_{_+} &0 &0  
	\end{array}
	\right)\,,\no 
\\
&&\mf{X}_L=\left(
	\begin{array}{cccc}
		0 & 0 &(\beta_1+i\beta_2) t_{_-}^*  &(\beta_3+i\beta_4) t_{_-}^*  \\
		0 & 0 &(\beta_3-i\beta_4 ) t_{_-}  &(-\beta_1+i\beta_2) t_{_-} \\
		(\beta_5+i\beta_6 ) t_{_-}^* &(-\beta_7+i\beta_8) t_{_-}^* &0&0\\
		(\beta_7+i\beta_8 )  t_{_-}  &(\beta_5-i\beta_6) t_{_-} &0 &0 
	\end{array}
	\right)\,, \no   \\
	&&\mf{Y}_L=\left(
	\begin{array}{cccc}
		0 & 0 &(-\beta_6 -i\beta_5) t_{_-} &( -\beta_8 -i\beta_7) t_{_-}^* \\
		0 & 0 &(-\beta_8 +i\beta_7 ) t_{_-}  &( \beta_6 -i\beta_5) t_{_-}^*  \\
		(\beta_2 +i\beta_1) t_{_-}^* & (-\beta_4 +i\beta_3) t_{_-} &0 &0  \\
		(\beta_4 +i\beta_3) t_{_-}^* & (\beta_2 -i\beta_1 ) t_{_-} &0  &0  
	\end{array}
	\right)
\eea
leads  to the fluctuation Lagrangian with 
 constant coefficients 
\bea \label{ferpw} &&
\mathcal{L}_{\rm F}=2\Big[\sum\limits_{i=1}^8 \left( \alpha_i \partial_-\alpha_i +\beta_i
 \partial_+\beta_i\right) 
-\mu \left( \alpha_1\alpha_2+\alpha_3\alpha_4+\alpha_5\alpha_6-\alpha_7\alpha_8-\beta_1\beta_2-\beta_3
\beta_4-\beta_5\beta_6+\beta_7\beta_8 \right) \no 
\\ &&\hspace{56pt}
+2 \kappa \left(  \alpha_1\beta_4 +\alpha_2\beta_3   -\alpha_3\beta_2 -\alpha_4\beta_1 -\alpha_5\beta_8
  +\alpha_6\beta_7 +\alpha_7\beta_6  - \alpha_8\beta_5    \right)
\Big] \,.
\eea
It  describes 8 fermionic  degrees of freedom  with
characteristic  frequencies equivalent (up to overall  shifts)\foot{These shifts reflect
particular redefinitions of the fermionic fields we have chosen.}
to $\sqrt{n^2 +\kappa^2}$. The  fermionic contribution to the one-loop 
partition function following from \rf{ferpw} is 
\be \label{ferpwd}
Z^{(1F)}=	\det \Big[ \partial_+^2 \partial_-^2+2 \partial_+ \partial_- \kappa ^2+\frac{1}{4} 
	 \mu ^2 \left(\partial_+^2 + \partial_-^2\right)+\frac{1}{16} \left(4 \kappa ^2-\mu 
	^2\right)^2 \Big] \,.
\ee 
By the same argument as in \eqref{detpwf3},\rf{shi} this 
 determinant  can be shown to be  equivalent to\foot{The
relation between determinants  implies rotation of the basis. 
This  extra rotation  leads, however, 
to non-constant coefficients in  the cubic and quartic terms in the fluctuation
 Lagrangian  and for this reason 
  keep using the rotation by $t_{_{\pm}}$ as  defined  above.}
  $[\det ( \partial_+ \partial_- + \kappa^2)]^2$.

 The final expression for the one-loop partition 
  function is thus the same as in string  theory \ci{ftt}. 
In the $\mu\to 0$ limit we  get, as in \rf{jok}, 
 the  familiar  result   \ci{ft1} (see \rf{kyi})
\bea
&&\G^{(1)} = {1 \ov 2}  \bar V_2 \int { d^2 q \ov (2 \pi)^2 }\big[ \ln (q^2 + 4 \kappa^2)
+ 2  \ln (q^2 + 2 \kappa^2)
 + 5 \ln q^2  - 8 \ln (q^2 +  \kappa^2) \big]\no\\ &&\ \ \ \ \ \ \ \ \ \ \ 
  = 2 \kappa^2 \bar V_2 \big(I[4]+ I[2] -2 I[1]\big) 
  =  { 1 \ov 2 \pi}  \aa_1   V_2 \ , \ \ \ \ \ \ \ \  \aa_1=-3 \ln 2 \ . \la{iok}
  \eea

\subsection{Two-loop contribution\label{sectwoloop}}

As in the \ads  case in the first approach discussed in 
 section 3.1 the two-loop computation uses 
the fluctuation Lagrangian in  \eqref{pw2}, \eqref{pw3}, \eqref{pw4} expanded 
near the above classical
 background \rf{gog}. Here one  treats  $4+4$ physical and 
 $6+6$ unphysical bosonic fluctuations on an equal footing. 
 Let us first consider the purely bosonic contributions   given by diagrams
  in Figure \ref{feybos}. 
 As discussed in Appendix \ref{detoneloop}, it is useful
 to make  $O(2)$ transformations of 
the unphysical fluctuations  (see \eqref{tc2c5} and 
\eqref{tc1c4})  to obtain four decoupled subsectors so that 
 the propagator takes a block-diagonal form.  
 If we  label the bosonic fluctuations as 
\be
\Phi _I =\{ \Phi_{A},\Phi_{S} \}, \ \ 
 \Phi_{A}=\{ a_1,c_2,c_3,a_2,c_5,c_6,a_3,a_4,c_1,c_4 \}  \,,\ \ ~~~~ \Phi_{S}
=\{ b_1,\dots,b_4,d_1,\dots,d_6\} \no
\ee
then  the euclidean-signature  propagator in the A-sector becomes 
\be \label{pads1}
 \triangle_{A}^{-1}(q)= \left( \ba{cccc}
 \M_1(q) & 0 & 0 & 0\\
 0 & \M_1(q) & 0 & 0\\
 0 & 0 & \M_2(q) & 0\\
 0 & 0 & 0 & \frac{1}{2 q^2}\ea
 \right)  \,,
\ee
where $\M_1(q)$ and $\M_2(q)$ are $3 \times 3$ matrices
(we again rescale $q$ by $\elk$, i.e. 
$\hat \kappa = { \kappa \ov \elk}, \ \hat \mu = { \mu \ov \elk}$) 
\be\label{pads2}\ba{c}
\M_1(q)= \frac{1}{D_1} \left(
\begin{array}{ccc}
 -\frac{4 \hat{\kappa}^4 -4 \hat{\kappa}^2  \hat \mu ^2+\hat \mu ^2 \left(q_{_+}
 ^{2}+\hat \mu ^2\right)}{4 \hat \mu ^2 \left(q_{_-}^2 +\hat \mu ^2\right) } &
  \frac{ \hat{\kappa} \sqrt{\hat{\kappa}^2 -\hat \mu ^2} \left(2 \hat{\kappa}^2
   -\hat \mu ^2\right)}{\sqrt{2} \hat \mu ^2 \left(q_{_+}^{2}+\hat \mu ^2\right)}
    & -\frac{ \hat{\kappa} \sqrt{ \hat{\kappa}^2 -\hat \mu ^2} q_{_+}}{\sqrt{2}
     \hat \mu  \left(q_{_+}^{2}+\hat \mu ^2\right) } \\
 \frac{ \hat{\kappa} \sqrt{\hat{\kappa}^2 -\hat \mu ^2} \left(2 -\hat \mu ^2\right)}
 {\sqrt{2} \hat \mu ^2 \left(q_{_+}^{2} +\hat \mu ^2\right)} & 
-\frac{4 \hat{\kappa}^2 \left(\hat{\kappa}^2 -\hat \mu ^2\right) \left(q_{_-}^{2} 
+\hat \mu ^2\right)+ \hat \mu ^2 \left(q^4 - \hat \mu ^4\right)}{2 \hat \mu ^2 
\left( q^4 +2 \hat \mu ^2 q^2-4 \hat \mu ^2 q_0^2+\hat \mu ^4\right)} & \frac{ 
\hat{\kappa}^2 q_{_+}  \left(q_{_-}^{2}+\hat \mu ^2\right) + i q^1 \hat \mu ^2
 \left(q^2- \hat \mu ^2\right)}{\hat \mu  \left(q^4+2 \hat \mu ^2 q^2-4  \hat 
 \mu ^2 q_0^2+ \hat \mu ^4\right)} \\
 \frac{ \hat{\kappa} \sqrt{ \hat{\kappa}^2-\hat \mu ^2} q_{_+}}{\sqrt{2} \hat \mu
   \left(q_{_+}^{2}+\hat \mu ^2\right)} & -\frac{\hat{\kappa}^2 q_{_+} \left(q_{_-}
   ^{2}+\hat \mu ^2\right) +i q^1 \hat \mu ^2 \left( q^2-\hat \mu ^2\right)}{ \hat
    \mu \left( q^4+2 \hat \mu ^2 q^2-4 \hat \mu ^2 q_0^2+ \hat \mu ^4\right)} & 
    \frac{q^4- \hat \mu ^4}{2 \left(q^4+2 \hat \mu ^2 q^2-4 \hat \mu ^2 q_0^2+ 
    \hat \mu ^4\right)}
\end{array}
\right) \,, \\
~\\
\M_2(q)= \frac{1}{D_2} \left(
\begin{array}{ccc}
 -\frac{q^2}{4 } & -\frac{\hat{\kappa}^2 q_{_-} + i q^1 \hat \mu ^2}{2 \hat \mu }
  & \frac{ \hat{\kappa} q_{_-}   \sqrt{\hat{\kappa}^2 -\hat \mu ^2}}{\sqrt{2}
   \hat \mu  } \\
 \frac{\hat{\kappa}^2 q_{_-} + i q^1 \hat \mu ^2}{2 \hat \mu  } & \frac{q_{_+}
  \left(q_{_+}^2 \hat \mu ^2 - 4 \hat{\kappa}^2 \left(\hat{\kappa}^2- \hat 
  \mu ^2\right)\right)}{4 q_{_-} \hat \mu ^2} & -\frac{ \sqrt{2} \hat \kappa 
   \sqrt{ \hat{\kappa}^2 -\mu ^2} \left( \hat{\kappa}^2 q_{_-} +i q_1 \hat
    \mu ^2\right)}{q_{_+}  \hat \mu ^2} \\
 -\frac{ \hat{\kappa} q_{_-}   \sqrt{\hat{\kappa}^2-\hat \mu ^2}}{\sqrt{2}
  \hat \mu } & -\frac{ \sqrt{2} \hat{\kappa}  \sqrt{\hat{\kappa}^2 -\hat 
  \mu ^2} \left(  \hat{\kappa}^2 q_{_-} +iq_1 \hat \mu ^2\right)}{q_{_+} 
  \hat \mu ^2} & \frac{q^4 \hat \mu ^2-4 \left(q^1 \hat \mu ^2-i \hat{\kappa}
  ^2 q_{_-} \right)^2}{2 q^2 \hat \mu ^2}
\end{array}
\right)  \,, \\
~\\
q_{_\pm}=q_0+i q_1\,,~~~~q^2=q_0^2+ q_1^2\,,~~~~D_1=q^2+2 \hat{\kappa}^2 -\hat 
\mu ^2\,,~~~~~D_2=q^4+4 \hat{\kappa}^2 q^2 -4 \hat \mu ^2 q_1{}^2 \,.
\ea\ee
The propagator in the S-sector is  (${\bf I}_n={\rm diag}(1,...,1)$)
\be \label{ps1}
\triangle_{S}^{-1} (q)= 
 {\rm diag}\left( -\frac{1}{4 \left(q^2+\hat\mu ^2\right)}  {\bf I}_4, -\frac{1}{2 q^2} 
{\bf I}_6 \right) 
\ . 
\ee
As the  two-loop computation for finite $\mu$ appears to be quite complicated, 
 we  consider only the limit $\mu \to 0$.
 This limit 
 (i.e. $\hat \mu \to 0,\ \hat \kappa \to 1$)
 can be smoothly taken once we simplify the integrands of the two-loop integrals.  
Below  we will summarize the results for the contributions 
of different types of two-loop  diagrams 
 found after  taking this limit.

\subsubsection{Bosonic 1PI contributions}

Plugging the bosonic fluctuation fields \eqref{etapr} 
 into ${\cal L}^{(3)}$ in \eqref{pw3} one finds the 
vertices for the sunset diagrams in Figure \ref{feybos}(a). 
Sunset diagrams  will be expressed in terms of the integrals \rf{wmm3}
\iff 
\be \label{wmmm}
	{\cal I}_{m_i^2m_j^2m^2_k}= \int \frac{d^2 \! q_i\,
	 d^2\! q_j\, d^2\! q_j}{(2\pi )^4}\frac{\mathcal{F}
	 (q_i,q_j,q_k)}{q_i^{n_i}q_j^{n_j}q_k^{n_k}(q^2_i+m_i^2)(q^2_j+m_j^2)(q^2_k+m_k^2)}\,,
\ee
\fi 
where  $\mathcal
{F}(q_i,q_j,q_k)$ is a polynomial function of momenta. 
In the A-sector, the vertices contained in the
 fluctuation Lagrangian are of the  three types. 
The first type includes the vertices $V_{A\, ijk}$ with $(i,
j,k)=\left(\{ 7,8,9\} ,\{ 1,2,3\} ,\{ 1,2,3\} \right)$ or $(i
,j,k)=\left(\{ 7,8,9\} ,\{ 4,5,6\} ,\{ 4,5,6\} \right)$. In the $\mu
 \to 0$ limit we have (see \rf{pads2})
  $D_1\to q^2+2$ and $D_2 \to (q^2+4)q^2$, so that  
 using  these vertices we obtain the integral ${\cal I}_{422}$ containing $I[4,2,2]$
and thus  the Catalan's constant   in \rf{jk}. 
The vertices $V_{A\, ijk}$ with $(i,j,k)=\left(\{ 7,8,9 \},\{ 
7,8,9\} ,\{ 7,8,9\} \right)$  
lead to  the integral ${\cal I}_{444}$. 
A nontrivial 
finite part of this integral  may contain   $I[4,4,4]$  but 
as 
in the corresponding string-theory computation
such term  does not actually appear. 
The third  type of vertices 
is $V_{A\, ijk}$ with $(i,j,k)=\left(10,\{ 1,2,3 \}
,\{ 4,5,6\} \right)$ yielding the integral ${\cal I}_{022}$. 

Explicitly, one finds for the resulting 
contribution to the two-loop effective action (cf. \rf{qw};  here $k=k_5$):
\bea \la{wq}
&& \Gamma^{(2)} =\frac{8 \pi}{k} V_2  \sum_n J_n \ , \ \\
  &&  J_{_{\rm boson~sunset}}={\cal I}_{422} + {\cal I}_{444} + {\cal I}_{022} \ , \
\ \ \ \ \ \ \ \ \ 
 {\cal I}_{422}=2 I[4,2,2]-\frac{1}{2} I[2]I[0]-\frac{3}{2} I[4]I[2]\,,\no \\ && 
 {\cal I}_{444}=-\frac{1}{4} I[4]I[0]-\frac{1}{4} I[4]I[4]\,,\ \ \ \ \ \ \ \ \ \ \  
 {\cal I}_{022}=-\frac{1}{2} I[2]I[2]\, , \label{bossun}
\eea
where we 
used 
the notation  in \eqref{uvandfin2}. 
The contribution from  the S-sector is trivial  as the corresponding part of the 
reduced theory solution  is  the  vacuum one.\foot{
Explicitly, one finds 
$$
\int \frac{d^2\! q_i \, d^2\! q_j}{(2\pi )^4}  \frac{q_i^+ 
q_j^--q_i^- q_j^+}{q^2_i q^2_j (q_i + q_j)^2 } 
 =0 
$$
due to obvious symmetry of the momentum-space integral 
under interchange $i \leftrightarrow j$.
}


The diagrams in Figure \ref{feybos}(b)  lead to momentum 
integrals of the type \rf{wmm3}.
In the A-sector  one finds:
\be\label{bosdd}\ba{l}
J_{_{\rm boson~double-bubble}}=	{\cal I}_{22} +  {\cal I}_{44}\ , \ \ \ \ \ 
{\cal I}_{22} =-\frac{1}{2} I[2]I[2]\,,\ \ \ \ \ \ \ \ \ 
	{\cal I}_{44}=-\frac{1}{4}I[4]I[0]-\frac{1}{4}I[4]I[4]\,.
\ea\ee
The S-sector again does not lead to a non-trivial contribution.


Summing up \eqref{bossun} and \eqref{bosdd} we find  the total  contribution of two-loop bosonic 
 1PI diagrams
\be \label{bostot}
J_{_{\rm boson~1PI}} =
	 2I[4,2,2]-\frac{1}{2}I[2]I[0]-\frac{1}{2}I[4]I[0]-I[2]I[2]-
	\frac{3}{2}I[4]I[2]-\frac{1}{2}I[4]I[4] \,, 
\ee
which contains the Catalan's constant term 
\be \label{catbos}
	I[4,2,2]= \frac{1}{16 \pi^2} \K\,.
\ee

\subsubsection{Fermionic 1PI contributions}


Let us  now consider the  1PI  diagrams with the fermionic propagators
 in Figure \ref{feyfer}.
As the fermionic fluctuations have mass  $\hat \kappa =1$ 
 the integrals arising 
from the fermionic sunset diagram 
are of the type  ${\cal I}_{m^2,1,1}$, where $m$ is a mass
 of the bosonic fluctuation (i.e. $m^2=4, 2,0$).  Explicitly, we find 
that the nonvanishing contributions are  
\bea \label{fsun} 
&&J_{_{\rm fermion-boson~sunset}}={\cal I}_{211} + {\cal I}_{411}+  {\cal I}_{011}\ , \ \ \ \  \ \ \ \ 
{\cal I}_{211}=-2 I[2,1,1]+I[1]I[1]-2I[2]I[1]\ , \no  \\   &&  
{\cal I}_{411}=I[4]I[1]-\frac{1}{2}I[1]I[1]\,, \ \ \ \ \ \ \ \ \ \ \ \
{\cal I}_{011}=9I[1][0]-\frac{9}{2}I[1]I[1]\,.
 \label{fersun} \eea
The fermionic double-bubble diagrams are expressed in terms of the 
integrals  ${\cal I}_{m^2,1}$,  where the $m$  is again the bosonic fluctuation mass:
\bea 
&&J_{_{\rm fermion-boson~double-bubble}} ={\cal I}_{41} + {\cal I}_{21} + {\cal I}_{01} \ , \no \\ 
&&\ \ \ \ \ \ \ \ \ \ \ \ {\cal I}_{01}=-9I[1]I[0]\,,\ \ \ \ 
{\cal I}_{21}=6I[2]I[1]\,,\ \ \ \ 
{\cal I}_{41}=I[1]I[0]+2I[4]I[1]\,.  \label{ferdd} 
\eea
Combining \eqref{fersun} and \eqref{ferdd} we get 
\be \label{fertot}
J_{_{\rm fermion~1PI}}
 = -2 I[2,1,1]+ I[1]I[0] -4 I[1]I[1] +4 I[2]I[1] +3 I[4]I[1] \,, 
\ee
which again contains the Catalan's constant term since 
\be
	 I[2,1,1]=  \frac{1}{8 \pi^2} \K\,.
\ee
Combining the bosonic \rf{bostot} and 
the fermionic \rf{fertot}   1PI contributions together we find 
\be \la{1pi}
J_{_{\rm 1PI}}
= - \frac{1}{8\pi^2} \K
- { 1 \ov 2}  \big(I[4] + I[2] - 2 I[1]\big) 
             \big(I[4] +  2 I[2] + I[0] - 4 I[1]\big)
\ . \ee
We observe that  as in the corresponding string theory computation \ci{rtt} 
the  fermionic  Catalan's constant contribution is  twice and  opposite
in sign  to the bosonic one.    Also, as in the  \ads reduced theory 
case \rf{jus}, the second term in \rf{1pi} is UV finite but IR divergent
and is proportional to  the same combination $I[4] + I[2] - 2 I[1]$
which appears in the one-loop  result \rf{iok}.

\subsubsection{Tadpole contributions  and  total result for the two-loop coefficient
 \label{secpwtad}}

The non-1PI diagrams relevant in the present case are shown 
 in Figure \ref{feytad}.\foot{To evaluate them we again use the prescription  
that momentum of the  intermediate line 
is set to zero only at the end of the computation.} We find for their contributions
(including  the $ { 1\ov 8}$  combinatorial factor) 
\bea
&& J_{_{\rm boson-boson~tadpole}}= {1 \ov 6} 
  \big(  I[2]I[0] +  I[4]I[0]  + 3  I[2] I[2]
  + 5 I[4]I[2]  + 2  I[4]I[4] \big) \,, \\
&&J_{_{\rm boson-fermion~tadpole}}
=  - { 1 \ov 3}\big( I[1]I[0]  + 6 I[2]I[1] + 5 I[4]I[1]  \big) \,, \\
&&
J_{_{\rm fermion-fermion~tadpole}}=  2  I[1] I[1] 
 \,.
\eea
The total tadpole contribution is thus 
\be\label{tadtot}
J_{\rm tadpole}=
 {1\ov 6}  \big(I[4] + I[2] - 2 I[1]\big)
          \big( 2I[4]  + { 3 } I[2]  +{  } I[0] - 6 I[1]  \big) 
  \ . \ee
Like the corresponding expression \rf{jas} found using first approach 
 in the  \ads  case,  this  coefficient  
is UV  finite but IR divergent and is  
proportional to the one-loop combination in \rf{iok}.

Combining together \rf{1pi} and \rf{tadtot} we find  the following 
expression for the coefficient in the two-loop effective action \rf{wq}
(cf. \rf{kyi})
\bea 
&& 
\sum_n J_n= J_{_{\rm 1PI}} + J_{\rm tadpole}= \bar J + \td J \ , \ \ \ \ \ 
\bar J = - {1 \ov 8 \pi^2} \K \ , \la{j}
\\
&& 
\td J= -\frac{1}{6} \big(I[4] + I[2] - 2 I[1]\big) \big( I[4]  + 
 3  I[2]  +2 I[0] - 6 I[1]  \big) \la{k}
\ . 
\eea
The resulting  two-loop coefficient thus 
contains, in addition to $\bar \aa_2=-\K$,  also 
$ \td \aa_2=8 \pi^2\td  J =    -\frac{5 }{4} (\ln 2)^2 -  \ln 2 \, \ln m_0 $  
which is  IR divergent. 

The close similarity  with the \ads   case discussed in  section 3
 suggests that  the  problem 
with non-cancellation of IR divergences is due to a subtlety in how the 
tadpole contribution  was computed.  In particular,  the analogy
with the \ads case  suggests that 
there should be no $I[4] I[4]$ term in the  tadpole contribution in 
\rf{tadtot}.
Indeed, the results in the  reduced \ads  and reduced \adss theories 
are  in direct agreement in what concerns $I[4] I[4]$ contributions
coming from the 1PI graphs:  this term enters \rf{1pi} with the same coefficient $-\ha$
as in \rf{jus} or \rf{kus}.

Accepting this  natural suggestion, the tadpole  term  \rf{tadtot}
should be replaced by\foot{Since the UV finiteness should  be preserved this    is 
effectively equivalent to replacing $I[4]$ in the second factor by $I[0]$, as in the 
$AdS_3\times S^3$ case.}
\bea
&& J'_{\rm tadpole}=
 {1\ov 6}  \big(I[4] + I[2] - 2 I[1]\big)\big(  { 3 } I[2]  +  3 I[0] - 6 I[1]  \big) \no 
\\	&& \ \ \ \ \  \ \ \ \  = 
 {1\ov 2}  \big(I[4] + I[2] - 2 I[1]\big)
          \big(   I[2]  +   I[0] - 2 I[1]  \big) 
  \ .\label{atot}  \eea
Then the sum of \rf{1pi} and \rf{atot} leads to 
$ \td \aa_2$ which is  IR finite and, as in the \ads case in \rf{kpq},  
is proportional to the square of the  one-loop coefficient: 
\be \la{ig}
\td J= -\frac{1}{2} \big(I[4] + I[2] - 2 I[1]\big)^2 
 \ , \ \ \ \ \ {\rm i.e.}\ \ \ \ \ \ \ \ \ 
  \td \aa_2= 8 \pi^2 \td J = - {1 \ov 4} (\aa_1)^2=- { 9 \ov 4} (\ln 2)^2   \ . 
\ee

\iff
For ${AdS}_5 \times { S}^5$ we have discussed the two-loop computation only 
based on the approach of using the PW identity, which corresponds to approach I in 
the $AdS_3 \times S^3$ case. 
It may be expected that we could carry out the two-loop computation in approach II
 by using the parameterization of $Sp(2,2) \times Sp(4)$ introduced in \cite{Fra} and 
 integrating out the gauge fields at the level of the classical Lagrangian, then 
 obtaining an extension of the model we used in section \ref{ads3s3sg}. 
However, such approach breaks down in the present case by some unclear reason; it 
does not reproduce the correct one-loop result, then we can not expect the two-loop 
result.  
So, we do not investigate the ${AdS}_5 \times { S}^5$ version of the formula
 \eqref{ads3conj}, while it is in principle possible to introduce the counterterms such
that the two-loop contribution takes the form \eqref{ads3conj}. 
 \fi

\section*{Acknowledgements}
We would like to thank B. Hoare and A. Rej for useful discussions. 
The  work of RR  was supported in part by the US National Science Foundation 
under grant PHY-0855356 and the A. P. Sloan Foundation.


\appendix

\section{Matrix superalgebra  
\label{appsupcos}}

In this Appendix we will 
briefly summarize some  notation  used in this paper (for details see
\cite{gt1,rtfin,hit,arf}). 
The superalgebra 
 $\mathfrak{su}(2,2|4)$ 
 is spanned by $8 \times 8$ supermatrices $\mathfrak{f}$
\begin{eqnarray}\label{f8by8}
&&	\mathfrak{f}= \left(
	\begin{array}{@{\,}cc@{\,}}
		\mathfrak{A} & \mathfrak{X} \\
		\mathfrak{Y} & \mathfrak{D}
	\end{array}	  
	\right) \,,\ \ \ \ \ \ 
	{\rm STr}\ \mathfrak{f}\equiv {\rm tr}\ \mathfrak{A}-{\rm tr}\ \mathfrak{D}=0\,,\\
	&& \mathfrak{A}^{\dagger}=-\hSigma \mathfrak{A} \hSigma \,,~~~~~
	\mathfrak{D}^{\dagger}= -\mathfrak{D}\,,~~~~~
	\mathfrak{Y}=-\mathfrak{X}^{\dagger}\hSigma \,\ , \ \ \ \ \ ~~
	\hSigma = {\rm diag} (1,1,-1,-1) \ , 
\end{eqnarray}
where the $4 \times 4$
  matrices $\mathfrak{A}$, $\mathfrak{D}$ are 
  Grassmann even and $\mathfrak{X}$, $\mathfrak{Y}$ are Grassmann odd. 
 $\mathfrak{A}$ belongs to $\mathfrak{u}(2,2)$ and 
 $\mathfrak{D}$ belongs to $\mathfrak{u}(4)$. 
 $\mathfrak{psu}(2,2|4)$ is the 
 quotient  of $\mathfrak{su}(2,2|4)$ over the  remaining 
  $\mathfrak{u}(1)$. 
 $\mathfrak{psu}(2,2|4)$  admits a $\mathbb{Z}_4$ decomposition: 
 \begin{eqnarray}
&&	\mathfrak{f}=\mathfrak{f}_0 \oplus \mathfrak{f}_{1}
\oplus \mathfrak{f}_2 \oplus \mathfrak{f}_3\,,\ \ \ \ \ \ \ \ \ \ \ \ \ \ \left[ \mathfrak{f}_r,
\mathfrak{f}_s \right] \subset \mathfrak{f}_{r+s~{\rm mod}~4} \,, \\
&& \mathfrak{f}_0=\frac{1}{2}\left(
	\begin{array}{@{\,}cc@{\,}}
		\mathfrak{A}-K\mathfrak{A}^{\rm t}K & 0 \\
		0 & \mathfrak{D}-K\mathfrak{D}^{\rm t}K
	\end{array}	  
	\right) \,, \nonumber \ \ \ \ \ \ 
	 \mathfrak{f}_1=
	 \frac{1}{2}\left(
	\begin{array}{@{\,}cc@{\,}}
		0 & \mathfrak{X}-iK\mathfrak{Y}^{\rm t}K \\
		\mathfrak{Y}+iK\mathfrak{X}^{\rm t}K & 0
	\end{array}	  
	\right) \,, \nonumber \\
	&& \mathfrak{f}_2=
	\frac{1}{2}\left(\begin{array}{@{\,}cc@{\,}}
		\mathfrak{A}+K\mathfrak{A}^{\rm t}K & 0 \\
		0 & \mathfrak{D}+K\mathfrak{D}^{\rm t}K
	\end{array}	  
	\right) \,, \ \ \ \ \ 
	\mathfrak{f}_3
	= \frac{1}{2}\left(
	\begin{array}{@{\,}cc@{\,}}
		0 & \mathfrak{X}+iK\mathfrak{Y}^{\rm t}K \\
		\mathfrak{Y}-iK\mathfrak{X}^{\rm t}K & 0
	\end{array}	  
	\right) \,, \nonumber \\
	&& K = \left(
	\begin{array}{@{\,}cccc@{\,}}
		0 & -1 & 0 & 0 \\
		1 & 0 & 0 & 0 \\
		0 & 0 & 0 & -1 \\
		0 & 0 & 1 & 0 
	\end{array}	  
	\right) \,.
\end{eqnarray}
  $\mathfrak{g}=\mathfrak{f}_0$  is  the algebra  of the 
group  $G=Sp(2,2) \times Sp(4)$ and
 $\mathfrak{p}=\mathfrak{f}_2$  as the  $F/G$ coset part of the algebra $\mathfrak{f}$. 
The element $T$ of the maximal Abelian subalgebra of $\mathfrak{p}$ 
\be \label{tdef}
T=\frac{i}{2}\trm{ diag}\left(1,\, 1,\, -1,\, -1,\, 1,\, 1,\, -1,\, -1\right)\, \ee
defines  a $\mbb{Z}_2$ decomposition ($r=0,1,2,3$)
\bea&& \mf{f}=\mf{f}^\parallel\oplus\mf{f}^\perp\,,\ \ \ \ \ 
\left[\mathfrak{f}^\perp,\,\mathfrak{f}^\perp\right]\subset\mathfrak{f}^\perp\,,\ \ \
\left[\mathfrak{f}^\perp,\,\mathfrak{f}^\parallel\right]\subset\mathfrak{f}^\parallel\,,
\ \ \
\left[\mathfrak{f}^\parallel,\,\mathfrak{f}^\parallel\right]\subset\mathfrak{f}^\perp\,,
\ \ \ \ \trm{STr}(\mf{f}^\parallel \mf{f}^\perp)=0\,, \\  &&
\mathfrak{f}^\parallel_r=-\left[T,\left[T,\mathfrak{
f}_r\right]\right]\,,\hspace{40pt} \mf{f}^\perp_r=-\{T,\{T,\mathfrak{f}_r\}\}\,,\ \ \
\ \ \ 
  \{\mathfrak{f}^\parallel_r, T\}=0, \ \ \  [ \mathfrak{f}^\perp_r, T]=0 \ .  \eea
We set   $\mf{h}=\mf{f}_0^\perp$, $\mf{m}=\mf{f}_0^\parallel$,
 so that 
 $\left[\mathfrak{h},\mathfrak{h}\right]\subset \mathfrak{h}, \ \
\left[\mathfrak{m},\mathfrak{m}\right]\subset \mathfrak{h}, \ \ 
\left[\mathfrak{m},\mathfrak{h}\right]\subset \mathfrak{m} 
$.
 $\mf{h}$  is the  algebra of the  subgroup  $H$ of $G$  which commutes with  $T$,  
$[ \mf{h},T]=0$.
For the specific choice of the matrices $\Sigma$, $K$ and $T$ 
 one can explicitly represent  the 
general elements of  $\mathfrak{m}$ and  $\mathfrak{h}$ as follows
(we use this  when discussing  the bosonic fluctuations in the reduced theory): 
\be
\ba{c}
\mathfrak{m}=\left( \ba{cc}
	\mathfrak{m}_{_A} & 0 \\
	0 & \mf{m}_{_S}
\ea \right) \,, \ \ \ \ \ \ \ \ 
\mf{m}_{_A} =\left(
	\begin{array}{cccc}
	0 & 0 & a_{1}+ia_{2}& a_{3}+ia_{4}\\
	0 & 0 & a_{3}-ia_{4}& -a_{1}+ia_{2}\\
	a_{1}-ia_{2}& a_{3}+ia_{4}& 0 & 0 \\
	a_{3}-ia_{4}& -a_{1}-ia_{2}& 0 & 0  
	\end{array}
	\right)\,, \\
\mf{m}_{_S}= \left(
	\begin{array}{cccc}
		0 & 0 & b_{1}+ib_{2}& b_{3}+ib_{4}\\
		0 & 0 & -b_{3}+ib_{4}& b_{1}-ib_{2}\\
		-b_{1}+ib_{2}& b_{3}+ib_{4}& 0 & 0 \\
		-b_{3}+ib_{4}& -b_{1}-ib_{2}& 0 & 0  
	\end{array}
	\right)\,, \\
\mf{h}=\left( \ba{cc}
	\mf{h}_{_A} & 0 \\
	0 & \mf{h}_{_S}
\ea \right) \,, \ \ \ \ \ \ \ \ 
\mf{h}_{_A}=\left(\ba{cccc}
i c_1 & c_2+i c_3&0&0
\\-h_c+i c_3&-i c_1&0&0
\\0&0&i c_4 & c_5+i c_6
\\0&0&-c_5+i c_6&-i c_4\ea\right)\, , \\
\mf{h}_{_S}=\left(\ba{cccc}
i d_{1} & d_2+i d_3&0&0
\\-d_2+i d_3&-i d_1&0&0
\\0&0&i d_4 & d_5+i d_6
\\0&0&-d_5+i d_6&-i d_4\ea\right)\, .
\ea
\ee
Fermionic fields of the reduced theory  take values in $\mf{f}^{\parallel}$, 
i.e. $\Psi_{_R} \in \mf{f}_1^{\parallel}$,  $\Psi_{_L} \in \mf{f}_3^{\parallel}$:
\be {\small \textstyle
\ba{c}
	\mf{f}_1^{\parallel}=\left(
	\begin{array}{cc}
		0& \mf{X}_1 \\
		\mf{Y}_1& 0 
	\end{array}
	\right) \,, ~~~~~~~~~~~~
	\mf{X}_1=\left(
	\begin{array}{cccc}
		 0 & 0 &\alpha_1+i\alpha_2 &\alpha_3+i\alpha_4 \\
		 0 & 0 &-\alpha_3+i\alpha_4 &\alpha_1-i\alpha_2 \\
		\alpha_5+i\alpha_6 &\alpha_7-i\alpha_8 &0&0\\
		\alpha_7+i\alpha_8 &-\alpha_5+i\alpha_6 &0 &0 \\
	\end{array}
	\right)\,, \\
	\mf{Y}_1=\left(
	\begin{array}{cccc}
		0 & 0 &-\alpha_6 -i\alpha_5 & -\alpha_8 -i\alpha_7 \\
		0 & 0 &\alpha_8 -i\alpha_7 & -\alpha_6 +i\alpha_5 \\
		\alpha_2 +i\alpha_1 & \alpha_4 -i\alpha_3 &0 &0  \\
		\alpha_4 +i\alpha_3 & -\alpha_2 +i\alpha_1&0 &0  
	\end{array}
	\right)\,,\\
	\mf{f}_3^{\parallel}=\left(
	\begin{array}{cc}
		0& \mf{X}_3  \\
		\mf{Y}_3& 0 
	\end{array}
	\right) \,,  ~~~~~~~~~~~~
	\mf{X}_3=\left(
	\begin{array}{cccc}
		0 & 0 &\beta_1+i\beta_2  &\beta_3+i\beta_4  \\
		0 & 0 &\beta_3-i\beta_4   &-\beta_1+i\beta_2 \\
		\beta_5+i\beta_6  &-\beta_7+i\beta_8  &0&0\\
		\beta_7+i\beta_8    &\beta_5-i\beta_6  &0 &0 
	\end{array}
	\right)\,,   \\ 
	\mf{Y}_3=\left(
	\begin{array}{cccc}
		0 & 0 &-\beta_6 -i\beta_5 & -\beta_8 -i\beta_7 \\
		0 & 0 &-\beta_8 +i\beta_7   & \beta_6 -i\beta_5  \\
		\beta_2 +i\beta_1 & -\beta_4 +i\beta_3 &0 &0  \\
		\beta_4 +i\beta_3 & \beta_2 -i\beta_1  &0  &0  
	\end{array}
	\right)\, . 
\ea }
\ee
The discussion of  $\mathfrak{psu}(1,1|2)$ superalgebra 
relevant for the \ads case   is similar. Here 
the matrix $\mathfrak{f}$ in \eqref{f8by8}
 is a $4 \times 4$ supermatrix with  $\mathfrak{A}$,
  $\mathfrak{B}$, $\mathfrak{X}$ and $\mathfrak{Y}$  being $2 \times 2$ matrices
  and 
\bea &&
		\Sigma =K= \trm{ diag}\left( 1,\,  -1\right)\ , 
		\ \ \ \ \ \ \ \ \
		T=\frac{i}{2}\trm{ diag}\left( 1,\,  -1,\, 1,\, -1 \right)\,,
\\  &&
\mathfrak{m}=\left( \ba{cc}
	\mathfrak{m}_{_A} & 0 \\
	0 & \mf{m}_{_S}
\ea \right) \,, \ \ \ \ 
\mf{m}_{_A} =\left(
\begin{array}{cc}
 0 & a_1+i a_2 \\
 a_1-i a_2 & 0
\end{array}
\right)\,, ~~~~~
\mf{m}_{_S}= \left(
\begin{array}{cc}
 0 &  b_1 +i b_2 \\
   b_1-i b_2 & 0
\end{array}
\right)\,, \no 
\\   &&
\mf{h}=\left( \ba{cc}
	\mf{h}_{_A} & 0 \\
	0 & \mf{h}_{_S}
\ea \right) \,, \ \ \ \ 
\mf{h}_{_A}=\left(
	\ba{cc}
	i c & 0 \\
	0& -i c
	\ea \right) \, , ~~~~~
\mf{h}_{_S}= \left(
	\ba{cc}
	i d & 0 \\
	0& -id
	\ea \right) \, .
\eea
The fermionic fields belong to  
\bea 
&&
	\mf{f}_1^{\parallel}=\left(
	\begin{array}{cc}
		0& \mf{X}_1 \\
		\mf{Y}_1& 0 
	\end{array}
	\right) ,\ \ \   \label{fxr33}
	\mathfrak{X}_1=\left(
\begin{array}{cc}
 0 & \alpha_1 +i \alpha_2  \\
  \alpha_3+i \alpha_4 & 0
\end{array}
	\right), ~~~
	\mathfrak{Y}_1=\left(
\begin{array}{cc}
 0 & -i \alpha_3-\alpha_4  \\
 i \alpha_1 +\alpha_2   & 0
\end{array}
	\right)\,\no 
\\ &&  
	\mf{f}_3^{\parallel}=\left(
	\begin{array}{cc}
		0& \mf{X}_3  \\
		\mf{Y}_3& 0 
	\end{array}
	\right) ,
\ \ \
	\mf{X}_3=\left(
\begin{array}{cc}
 0 & \beta_1 +i \beta_2 \\
 \beta_3+i \beta_4 & 0
\end{array}
	\right),   ~~~
	\mf{Y}_3=\left(
\begin{array}{cc}
 0 & -i \beta_3-\beta_4 \\
 i \beta_1+\beta_2 & 0
\end{array}
	\right)  \label{fyl33}
\eea

\section{Two-loop computation in $AdS_3 \times S^3$ 
superstring theory\label{ads3s3ost}}

Here we shall discuss the  computation of the two-loop  correction 
to the energy of long folded string  spinning in  ${ AdS}_3$
in critical  $AdS_3 \times S^3 \times T^4$   superstring theory. 
In this case it is sufficient to consider just 
the \ads supercoset theory   as extra massless modes  can be decoupled  \ci{bsz}.
The  calculation
is very similar to the 
one in the \adss case  \ci{rtt,rt1}. 
The efficient approach is to map  the infinite spin limit of 
the folded string  solution 
to the Poincar\'e coordinates, where it is a critical point
 of the Euclidean action equivalent to the null-cusp solution. 
We shall follow the light-cone gauge approach developed in \ci{grrtv}.
The strategy will be to compute the two-loop partition function on the corresponding
classical background. 

As was shown in \cite{MTT,MT2000},
the \ads   Green-Schwarz superstring  action  in the $AdS$ 
 light-cone gauge may be obtained 
by a simple truncation of the ${AdS}_5\times { S}^5$ light-cone gauge 
action: one is to  ignore 
the two boundary coordinates transverse to the light-cone, set to zero 
 two of the six transverse 
coordinates and reduce the number of components of fermions from 4 to 2. 
Starting from the action in \cite{MTT,MT2000}, setting $p^+=1$ and 
Wick-rotating to the  euclidean 
worldsheet signature by  $\sigma\rightarrow i\sigma$ leads to
(see, e.g., \ci{grrtv} for notation)
\by
&& S_E  = {\sql \ov 4 \pi}  \int d \tau \int^\infty_0 d \sigma \  L_E   \ , 
\label{ae} \\
&&
 L_E  = \big[ \dot z^M  +
        \frac{{\rm i}}{z^{2}} z_N \eta_i  (\rho^{MN}){}^i{}_j \eta^j   \big]^2
        + \   \frac{1}{z^{4}}  z'^Mz'^M  
+  {\rm i}  \big(\theta^i \dot{\theta}_i+\eta^i\dot{\eta}_i 
        +\theta_i \dot{\theta}^i +\eta_i\dot{\eta}^i \big) 
	 \no \\  && \ \ \ \ \ \ \ \ -  \  \frac{1}{ z^{2}} 
	  (\eta^i\eta_i)^2
+ \   2 {\rm i}  \Big[ \, \frac{1}{z^{3}}z^M \eta^i (\rho^M)_{ij} \theta'^j +
                                     \frac{1}{z^{3}}z^M \eta_i (\rho^\dagger_M)
				     ^{ij}\theta'_j \Big]  \ . 
\label{lae}
\ey
The form of the classical solution is the same as in the \adss  case: 
\by
{z}=\sqrt{\frac{\tau}{\sigma}}
\ , \ \ \ \ \ 
x^+ =  \tau \ , \ \ \ \ \ \ \ \ \ \ 
 x^-  = - { 1 \over 2\sigma} \ , \ \ \ \ \ \ \ \ \ 
 x^+ x^- = - \frac{1}{2} z^2  
\ , \ey
where the \ads 
 metric  is $ds^2 = { 1 \ov z^2} ( dx^+ dx^-  + dz_M dz_M$),\  $M=(a,4), \
 a=1,2,3.$
The fluctuations around the classical solution are defined as
\by
\label{flu}
&& z=\sqrt{\frac{\tau}{\sigma}}\ {\tilde z} \ , \ \ \ \ \ \ \ \ 
{\tilde z} = e^{\tilde \phi}= 1 + \tilde \phi  +\dots~,\ \ \  
 z^M=\sqrt{\frac{\tau}{\sigma}}\ {\tilde z}^M \ , \ \ \ \ 
{\tilde z}^M = e^{\tilde \phi} \tilde u^M \ ,  \\
&&
{\tilde u}{}^{a}=  \frac{y^{a}}{1+\frac{1}{4}y^2}~, \ \ \ \ 
{\tilde u}{}^{4} =  \frac{1-\frac{1}{4}y^2}{1+\frac{1}{4}y^2}  \ , \ \ \ \ \ \ \ \ \
~~~~ y^2\equiv \sum_{a=1}^3 (y^a)^2\ , 
\label{u} \\
&&
\theta=\frac{1}{\sqrt{\sigma}}{\tilde\theta}
~,~~~~~~
\eta=\frac{1}{\sqrt{\sigma}}{\tilde\eta}\ . 
\label{xx}
\ey
It is useful to do a  further redefinition of the worldsheet coordinates 
$(\tau,\sigma) 
\to (t,s)$ (we will 
denote by $(p_0,p_1)$ the corresponding two-dimensional momenta, i.e. 
$(p_0,p_1)=-{\rm i} (\partial_t,\partial_s)$)
\be
\label{oo} 
t=\frac{1}{2}\ln \tau~,~~~~~~s=\frac{1}{2}\ln \sigma~,~~~~~~~~~~~dt ds=\frac{1}{4}
\frac{d\tau d\sigma}{\tau\sigma}
~,~~~~~~~\tau\partial_\tau=2\partial_t
~,~~~~~~~~\sigma\partial_\sigma=2\partial_s\ . 
\ee
It leads then to the following euclidean action (\ref{ae}), (\ref{lae}):
\bea\label{action}
S_E &=&  {\sql \ov 4 \pi} \int dt \int^\infty_{-\infty} ds\ {\cal L}\ ,\\
\label{Lagrangian}
{\cal L}  &=&\big[ \partial_t {\tilde z}^M + {\tilde z}^M  +
\frac{2{\rm i} }{{\tilde z}^2} 
{\tilde \eta}_i  (\rho^{MN}){}^i{}_j {\tilde \eta}^j  {\tilde z}_N \big]^2 
+ \  \frac{1}{{\tilde z}^{4}} \big(\partial_s{\tilde z}^M - {\tilde z}^M \big)^2
 \cr
&+&  2{\rm i} 
({\tilde \theta}^i \partial_t{\tilde \theta}_i
+{\tilde \eta}^i\partial_t{\tilde \eta}_i + {\tilde \theta}_i
\partial_t{\tilde \theta}^i
+{\tilde \eta}_i\partial_t{\tilde \eta}^i)
-  \ \frac{4}{{\tilde z}^{2}} ({\tilde \eta}^2)^2 
\cr
&+&  4{\rm i}\ \Bigl[\ \frac{1}{{\tilde z}^{3}}{\tilde \eta}^i (\rho^M)_{ij} {\tilde z}^M
(\partial_s{\tilde \theta}^j -{\tilde \theta}^j )
           +
\frac{1}{{\tilde z}^{3}}{\tilde \eta}_i (\rho^\dagger_M)^{ij} {\tilde z}^M
(\partial_s {\tilde \theta}_j - {\tilde \theta}_j )\Bigr] \ . 
\eea 
The normalization of the worldsheet coordinates was chosen so
 that the masses 
of the quadratic fluctuations 
reproduce the masses of the fluctuations around the closed 
folded string, i.e.~the spectrum is given by one 
boson with $m^2=4$, three massless bosons
 and four  fermions with $m^2=1$. This may be obtained 
from the spectrum of fluctuations in the 
${AdS}_5\times { S}^5$  case by truncating away 
 two transverse ${AdS}_5$ fluctuations, 
two ${ S}^5$ fluctuations and half of the fermions.
With this normalization, the effective action is related to the cusp anomaly 
$\f(\lambda)$ as \cite{grrtv}
\bea
&&\Gamma = \frac{1}{2\pi }\; \hat f
(\lambda)\; V_2 \ ,\ \ \ \ \ \ \ \ \ \ \ \ V_2= \int dt\int_{-\elk/2}^{\elk/2} ds
~,~~~~~~~~\elk=2\pi \kappa = 2\ln \S \ , \\ 
&&\hat f
={\sql} + f \ , \ \ \ \ \ \ \ \ f=\aa_1
+\frac{1}{\sql}\aa_2+{\cal O}({ 1 \ov (\sql)^2})\, .
\eea
 Evaluating the one-loop 
effective action 
implies that the one-loop coefficient in the cusp anomaly is given by 
\by
a_1 =-{2\ln 2} \,. 
\ey
 Before quoting 
the result for the two-loop correction it  is  
instructive to discuss the  expected differences 
compared to the known 
 ${AdS}_5\times { S}^5$ result  which are due to the absence of the two 
 bosonic fluctuations with $m^2=2$.  
The terms that are not  given by  products of factors 
of lower transcendentality 
arise solely from 3-propagator integrals (such as $I[2,2,4]= { 1 \ov 16 \pi^2} 
 {\K}$ 
in $AdS_5$ case). From the 
bosonic diagrams we may expect $I[4,4,4]$, $I[4,4,0]$, $I[4,0,0]$ and $I[0,0,0]$
 while the fermionic 
diagrams may contribute $I[1,1,4]$ and $I[1,1,0]$. 
The integral $I[4,4,4]$ is generated by the terms in the Lagrangian that 
depend only 
on the ``radial'' ${AdS}_3$ coordinate $z$.  Such 
 terms are the same as in the \adss case; 
since there this contribution  canceled  out, it should 
  not  appear here either.
 A similar reasoning can be 
used to rule out all other  3-propagator integral  contributions. 
 Therefore,  we should expect 
that the two-loop effective
action should be  given only by a
 sum  of products $I[m^2_i] I[m^2_j]$ 
of one-loop integrals.
Such products all canceled out 
in the \adss  case, and the same should  happen here too. 

Indeed, a direct calculation based on \rf{action}
shows that the relevant two-loop  Feynman diagrams 
 produce the
 following contributions 
\by
\Gamma^{(2)} &=& { 4 \pi \ov \sqrt \lambda}    V_2
 \sum_n J_n\ , \\
J_{_{\text{boson\;sunset}}}&=&\frac{1}{2}I[4]I[4] \ , 
\ \ \ \ \ \ \ \ \ 
J_{_{\text{boson\; double-bubble}}}=-\frac{1}{2} I[4]I[4] \ , 
\\
J_{_{\text{fermion-boson\;sunset}}} &=& \frac{1}{4}\left(-3I[0]I[1] + 2 I[1]I[1] + 
4 I[1]I[4]\right)\ , 
\\
J_{_{\text{fermion-boson double-bubble}}}&=&-\frac{1}{4}\left(-3I[0]I[1] 
+ 4 I[1]I[4]\right)\ , 
\\
J_{_{\text{tadpole}}}&=& -\frac{1}{2}I[1]I[1]  \, .
\ey
As a result, the 
sum of all contributions  is not only UV and IR finite but also 
vanishes,\foot{Note that
the bosonic contribution vanishes separately; the fermionic term vanish
 only after both the 
1PI and non-1PI contributions are combined together.}
 $\sum_n J_n=0$,
 i.e.  in contrast to the \adss  superstring case  where
$ a_2 = - \K$,  in the \ads   case we find that 
\by
a_2=0\,.\la{nun}
\ey 
It would be interesting to reproduce this string theory result from 
the  asymptotic Bethe ansatz conjectured in \ci{bsz}.

\section{Comments  on one-loop computation 
 in section 4  \label{detoneloop}}

The  quadratic fluctuation Lagrangian in section 4.1 
looks different from the  corresponding one in \adss  string theory but the two lead
to equivalent sets of  characteristic frequencies and the  one-loop determinants.  
Here  we shall comment on the structure 
of subsectors   of the bosonic fluctuation  Lagrangian \rf{adstotal}.
Let us start with  ${\cal L}^{(2)}_{1}$ in \rf{adstotal} 
 containing $a_1$ and $a_2$.
Integrating out $c_2$,$c_3$,$c_5$ and $c_6$
gives\foot{The resulting determinant of the operator 
${\cal O}= \partial_+^2 + M_2^2$ is equivalent to the (square of) 
 massless operator 
 determinant.
}
\be \label{la1a2reap}
 \td {\cal L}^{(2)}_{1} = 
  2 \sum _{i=1,2} \Big[  \partial_+ a_i \partial_- a_i -
  \left(2 \kappa ^2-\mu ^2\right) a_i^2+ 4 M_1^2  a_i \frac{\partial_+ \partial_-
   + 2 \kappa ^2-\mu ^2}{ \partial_+^2 +M_2^2} a_i \Big] \,.
\ee
This looks   different from the fluctuation Lagrangian found 
from the corresponding string action 
(and found also in the reduced theory   by taking  the $\mu\to 0$ limit 
in the ``mixed'' gauge where the physical and unphysical modes are decoupled \cite{hit})
\be \label{la1a2stap}
 {\cal L}_1 =  2 \sum _{i=1,2} \big[  \partial_+ a_i \partial_- a_i - 
 \left(2 \kappa ^2-\mu ^2\right) a_i^2 \big] \,,
\ee
but the two are closely related  as one can factorise the operator 
$\partial_+ \partial_-  + 2 \kappa ^2-\mu ^2$ in \rf{la1a2reap}.

The   Lagrangians ${\cal L}^{(2)}_{1}$ in \rf{adstotal}  and $ {\cal L}_1$
   are, in fact,  related 
 by a nonlocal transformation. To see  this, it is useful to perform 
   the following  $O(2)$ 
rotations, 
\be  \label{tc2c5} \ba{c}
	c_2 \to \frac{1}{\sqrt{2}} \left( c_2 - c_6 \right) \,, \ \ \ \
	c_3 \to \frac{1}{\sqrt{2}} \left( c_3 + c_5 \right) \,, \ \ \ \
	c_5 \to \frac{1}{\sqrt{2}} \left( c_2 + c_6 \right) \,, \ \
	c_6 \to \frac{1}{\sqrt{2}} \left( c_3 - c_5 \right) \,. 
\ea \ee
Then  ${\cal L}^{(2)}_{1}$  splits into smaller  subsectors.
 One  contains $a_1$, $c_2$ and $c_3$, 
\by 
&& {\cal L}^{(2)}_{a_1} =  2 \big[ \partial_+ a_1 \partial_- a_1 - 
 \left(2 \kappa ^2-\mu 
^2\right) a_1^2\big]
   - 4 M_1 \left( \mu c_2 + \partial_- c_3  \right) a_1 \nonumber \\ && 
\hspace{20pt} 
- \sum _{j=2,3} \big[ \partial_+ c_j \partial_- c_j +(2 \kappa ^2-\mu ^2)
 c_j^2 \big] 
-2 \left( \mu \partial_+ c_3 + M_2 \partial_- c_3  \right) c_2 
 \,. \label{la1rot}
\ey
Another contains  $a_2$, $c_5$ and $c_6$  with a similar Lagrangian.
To decouple  $a_1$  from  $c_2$, $c_3$ we may  apply 
the nonlocal transformation 
\by
	&& a_1 \to a_1  -\frac{\sqrt{2} \kappa  \sqrt{\kappa ^2-\mu 
	^2}}{\partial _+ \partial _- +2 \kappa ^2-\mu ^2} c_2  
	- \frac{ \sqrt{2} \kappa \sqrt{\kappa^2-\mu^2} \partial _-
	 (\partial_- ^2 + \mu^2 )}{\mu (\partial _+^2 \partial _-^2 
	 - \mu ^4) } c_3\,, \nonumber \\
&& c_2 \to c_2+ \frac{\partial _+ \mu ^2 \left(\partial _+ \partial _-
+2 \kappa ^2-\mu ^2\right)+\partial _- \left(\left(2 \kappa ^2- \mu
 ^2\right)\partial _+ \partial _- +\mu ^4\right)}{\mu  \left(\partial
  _+^2 \partial _-^2-\mu ^4\right)} c_3 \,, \label{nona1}
\ey
leading to 
\by 
&& {\cal L}^{(2)}_{a_1} =  2\big[  \partial_+ a_1 \partial_- a_1 -  \left(2 \kappa
 ^2-\mu ^2\right) a_1^2 \big]  \no \\  && \hspace{20pt} 
+ 2 c_2 \frac{ \partial _+^2 \partial _-^2- \mu ^4}{ \partial _+ \partial
 _-  +2 \kappa ^2-\mu ^2} c_2  
+  c_3 \frac{ \left( \partial _+ \partial _-  +2 \kappa ^2-\mu ^2\right) 
\left(\partial _-^2+\mu ^2\right) \left(\partial _+ ^2+\mu ^2\right)}
{\partial _+ ^2 \partial _-^2 -\mu ^4} c_3  \,. \label{a1en}
\ey
The physical part of this Lagrangian is the same as \eqref{la1a2stap}.
The product of  determinants resulting  from integrating out 
$c_2$ and $c_3$ contains  only trivial massless factors. 
The same is true  in the  $a_2$,$c_5$,$c_6$ sector.

Similar observations apply in  the sectors 
 containing $a_3$ and $a_4$ described by the 
Lagrangian ${\cal L}^{(2)}_{2} $ in \rf{adstotal}. 
Integrating out $c_3,c_4$ directly leads to 
\be \label{adsl22}
\td {\cal L}_2 ^{(2)}=  2 \sum _{i=3,4}  \partial_+ a_i \partial_- a_i 
+ 4 \left( \mu \partial_+a_4 + M_2 \partial_-a_4 \right) a_3
+8 M_1^2 a_3 \frac{\partial_-}{\partial_+}a_3
 \,,
\ee
which looks different from the string theory counterpart 
\be \label{adsl2st}
	{\cal L}_{2}=  2 \sum _{i=3,4}  \partial_+ a_i \partial_- a_i 
	 +  4 \left(\kappa ^2-\mu ^2\right) a_3^2
	+4 \mu \left( \partial_+a_3+\partial_-a_3 \right) a_4\,. 
\ee
To find a transformation  between ${\cal L}^{(2)}_{2} $
and  \eqref{adsl2st}  let us  apply an $O(2)$ rotation 
\be \label{tc1c4} \ba{c}
	c_1 \to \frac{1}{\sqrt{2}} \left( c_1 + c_4 \right) \,, \ \ \ \ \ \ \ \ \ \ \ 
	c_4 \to \frac{1}{\sqrt{2}} \left( c_1 - c_4 \right) \,, 
\ea \ee
and  the following redefinition 
\be \label{nona3} \ba{c}
	c_1\to c_1+ 2 \sqrt{2} \frac{\kappa \sqrt{\kappa^2 -\mu^2}}{\mu} 
	\frac{1}{\partial_+} a_3 \, ,\ \ \ \ \ \ \ \ \ \ \ 
	a_4\to -a_4- 2 \frac{\kappa^2 -\mu^2}{\mu} \frac{1}{\partial_+}
	 a_3\,. 
\ea\ee
Then we get 
\by \label{a3a4en}
 &&  {\cal L}^{(2)}_{a_3,a_4} =  2 \sum _{i=3,4}  \partial_+ a_i \partial_- a_i  + 
  4 \left(\kappa ^2-\mu ^2\right) a_3^2
	+4 \mu \left( \partial_+a_3+\partial_-a_3 \right) a_4 - \sum _{j=1,4} 
	 \partial_+ c_j \partial_- c_j \,,
\ey
where the  physical part is  the same as in \eqref{adsl2st}.

\iff 
 
In this subsector, the  $\mu \to 0$ limit is not well-defined at the 
level of the Lagrangian \eqref{adsl2}, but the limit can be taken only 
after deriving the characteristic frequencies. Also we find, in the
 Lagrangian after integrating out the unphysical fields, \eqref{adsl22},
  that the nonlocal term becomes singular by taking this limit. These 
  are the same observation as in the subsector containing $a_1$ and $a_2$. 
The argument in this Appendix shows that there is no simple relation between
 the fluctuation Lagrangians in the reduced theory and the original theory. 
This is largely different from the result obtained in the decoupling gauge 
in \cite{hts,iwa}. 
Another important lesson is that the $\mu \to 0$ limit can not be taken at 
the level of the Lagrangian, requiring that we should start the two-loop
 computation with $\mu$ arbitrary.  
\fi

\section{The two-loop computation in the vacuum case }

In  the main part of this paper we studied two-loop  corrections near 
the folded  string with large spin $S$ in $AdS_3$    taking the limit $\mu\to 0$
in which the   angular momentum in $S^5$ vanishes. 
Here we shall   check that the two-loop correction vanishes  in the opposite limit of the trivial 
reduced theory solution corresponding to the  BMN vacuum, i.e. in the  case when 
\be  \la{li}  \kappa \to \mu \ ,\ \ \ \ \ \ \ \ \ \ \ \ \elk\to 0 
 \  . \ee 
In this  case it  is  useful to define the ``S'' part of  fluctuation fields 
with an additional rescaling by $\ww= e^{i \mu \tau}$ as follows 
(cf.  \rf{etapr} ) 
\be\label{etaprb}\ba{c}
	\eta^\parallel _S=\left(
\begin{array}{cccc}
 0 & 0 & b_1+i b_2 & (b_3+i b_4) \ww \\
 0 & 0 & (-b_3+i b_4) \ww^* & b_1-i b_2 \\
 -b_1+i b_2 & (b_3+i b_4)\ww & 0 & 0 \\
 (-b_3+i b_4)\ww^* & -b_1-i b_2 & 0 & 0
\end{array}
\right) \,, \\ 
\eta^\perp _S=\left(
\begin{array}{cccc}
 i d_1 & (d_2+i d_3) \ww & 0 & 0 \\
 (-d_2+i d_3 ) \ww^* & -i d_1 & 0 & 0 \\
 0 & 0 & i d_4 & (d_5+i d_6 )\ww \\
 0 & 0 & (-d_5+i d_6 )\ww^* & -i d_4
\end{array}
\right) \,.
\ea\ee
Also, the  component fields of the fermionic fluctuations are  to be defined as 
(cf. \rf{fxr})
\begin{gather} \label{fxrb}
	\mathfrak{X}_R=\left(
	\begin{array}{cccc}
		 0 & 0 &(\alpha_1+i\alpha_2) t_{_{1+}} &(\alpha_3+i\alpha_4)t_{_{2+}} \\
		 0 & 0 &(-\alpha_3+i\alpha_4) t_{_{2+}}^* &(\alpha_1-i\alpha_2) t_{_{1+}}^* \\
		(\alpha_5+i\alpha_6) t_{_{1+}} &(\alpha_7-i\alpha_8) t_{_{2+}}&0&0\\
		(\alpha_7+i\alpha_8) t_{_{2+}}^* &(-\alpha_5+i\alpha_6) t_{_{1+}}^* &0 &0 \\
	\end{array}
	\right)\,, \\
	\mathfrak{Y}_R=\left(
	\begin{array}{cccc}
		0 & 0 &(-\alpha_6 -i\alpha_5) t_{_{1+}} ^* & (-\alpha_8 -i\alpha_7) t_{_{2+}}  \\
		0 & 0 &(\alpha_8 -i\alpha_7) t_{_{2+}}^* & (-\alpha_6 +i\alpha_5) t_{_{1+}}  \\
		(\alpha_2 +i\alpha_1) t_{_{1+}} ^* & (\alpha_4 -i\alpha_3 ) t_{_{2+}}  &0 &0  \\
		(\alpha_4 +i\alpha_3) t_{_{2+}}^*  & (-\alpha_2 +i\alpha_1)t_{_{1+}}  &0 &0  
	\end{array}
	\right)\,,\\ 
	\mf{X}_L=\left(
	\begin{array}{cccc}
		0 & 0 &(\beta_1+i\beta_2) t_{_{2-}}^*  &(\beta_3+i\beta_4) t_{_{1-}}^*  \\
		0 & 0 &(\beta_3-i\beta_4 )t_{_{1-}} &(-\beta_1+i\beta_2)t_{_{2-}} \\
		(\beta_5+i\beta_6 )t_{_{2-}}^*  &(-\beta_7+i\beta_8) t_{_{1-}}^* &0&0\\
		(\beta_7+i\beta_8 ) t_{_{1-}}  &(\beta_5-i\beta_6)t_{_{2-}}  &0 &0 
	\end{array}
	\right)\,,   \\
	\mf{Y}_L=\left(
	\begin{array}{cccc}
		0 & 0 &(-\beta_6 -i\beta_5) t_{_{2-}} &( -\beta_8 -i\beta_7)t_{_{1-}}^* \\
		0 & 0 &(-\beta_8 +i\beta_7 )t_{_{1-}}  &( \beta_6 -i\beta_5)t_{_{2-}}^*  \\
		(\beta_2 +i\beta_1) t_{_{2-}} & (-\beta_4 +i\beta_3) t_{_{1-}}^* &0 &0  \\
		(\beta_4 +i\beta_3) t_{_{1-}} & (\beta_2 -i\beta_1 ) t_{_{2-}}^* &0  &0  
	\end{array}
	\right)\,,  \label{fylb}
\end{gather}
where 
\be
	t_{_{1\pm}}=e ^{i \frac{\elk^2 (\tau  \pm \sigma )}{2 \mu }} \,,~~~~~~~~~~~~~
	t_{_{2\pm}}=e ^{i\frac{(\elk ^2+2\mu ^2) \tau  \pm \elk^2 \sigma}{2 \mu } } \,.
\ee
Taking  the  limit \rf{li} in the two-loop diagrams one finds cancellations 
between $A$ and $S$   sectors in each type of diagrams 
leading to  the vanishing  two-loop
correction. 

One can also check this cancellation directly,  by  expanding near the reduced theory 
counterpart of the    BMN vacuum  
\be
	g_0={\bf I}_{8\times 8} \,,\ \ \ \ \ \ \ \ \  A_\pm =0 \ . 
\ee
In this case the $\tau, \sigma$-dependent rescalings of fluctuations 
are not needed and 2d Lorentz invariance of the perturbation theory is manifest. 
One then finds for the  individual diagram 
 contributions to the coefficient in the two-loop effective action\foot{As
  above, here $A$ and $S$ stand for contributions from the fluctuations 
  corresponding to 
  reduced theory counterparts of the $AdS_5$ and $S^5$ sectors.}
\bea
&& {\rm bosonic\ sunset}: \ \ \ \ \ \ \ \ \ \ \ \ \ \ \ \ \ \ \ \ \ \ 	J_{A}=-J_{{ S}}=-\frac{3}{2} I[1] I[1]\,,
\no\\ 
&& {\rm bosonic\ double-bubble}: \ \ \ \ \ \ \ \  \  	J_{{A}}=-~J_{{ S}}=-\frac{1}{2} I[1]I[1]\,,
\no\\ &&
{\rm fermionic\ sunset}: \ \ \  \ \ \ \ \ \ \ \ \ \ 	\ \ \ \ \ \ \ 	J_{{A}}=-J_{{ S}}=-6I[0]I[1] + 3[1]I[1] \,,
\no\\ && 
{\rm fermionic\ double-bubble}: \ \ \ \ \ \ \   		J_{{A}}=-J_{{ S}}=-6 I[0]I[1] -4 I[1]I[1]\,,
\no\\   &&
{\rm tadpole}: \ \ \ \ \ \ \ \ \ \ \ \ \ \ \ 	\ \ \ \ \ \ \ \ \ \  \ \ \ 	\ \ \	J_{{A}}=-J_{{ S}}=0 \,.
\eea
We conclude again that the  sum of the $A$ and $S$   sector contributions vanishes.

\end{document}